\documentclass[aps,prb,groupaddress,twocolumn,showpacs,longbibliography]{revtex4-1}
\usepackage{amsmath}
\usepackage{amsthm, amssymb} 
\usepackage{bbold}                       
\usepackage{graphicx}
\usepackage{color}
\usepackage{times, verbatim}      

\usepackage{hyperref}
\hypersetup{       colorlinks=true, }

\begin{document}

\title{Landau level quantization for massless Dirac fermions in the spherical geometry:  Graphene fractional quantum Hall effect on the Haldane sphere }

\author{Michael Arciniaga}
\affiliation{Department of Physics \& Astronomy, California State University Long Beach,  Long Beach, California 90840, USA}
\author{Michael R. Peterson}
\affiliation{Department of Physics \& Astronomy, California State University Long Beach,  Long Beach, California 90840, USA}

\begin{abstract}
We derive the  single-particle eigenenergies and eigenfunctions for massless Dirac fermions confined to the surface of a sphere in the presence of a magnetic monopole, i.e., we solve the Landau level problem for electrons in graphene on the Haldane sphere.  With the single-particle eigenfunctions and eigenenergies we calculate the Haldane pseudopotentials for the Coulomb interaction in the second Landau  level  and calculate the effective pseudopotentials characterizing an effective Landau level mixing Hamiltonian entirely in the spherical geometry to be used in theoretical studies of the fractional quantum Hall effect in graphene.   Our treatment is  analogous to the formalism in the planar geometry and reduces to the planar results in the thermodynamic limit.  
\end{abstract}

\date{\today}


\maketitle

\section{Introduction}

The fractional quantum Hall effect (FQHE) provides a well-established experimental manifestation of a strongly correlated electron system supporting topologically ordered ground states. When quasi-two-dimensional electrons are placed in a strong perpendicular magnetic field of strength $B$ (tens of teslas) at very low temperatures ($T < 1$ K) such that the electron filling factor $\nu=2\pi l_B^2\rho$ is a rational fraction ($l_B=\sqrt{\hbar c/ eB}$ is the magnetic length and $\rho$ is the two-dimensional electron density) the kinetic energy is quenched (macroscopically degenerate Landau levels form), the low-energy physics is dominated by the electron-electron interaction, and an incompressible topological ordered quantum fluid forms~\cite{dasSarma1996,jainCF,tsui1982}. The experimental phenomena of the FQHE is the observation of a plateau in the Hall resistance $R_{xy} = h/fe^2$ along with a vanishing of the longitudinal resistance $R_{xx}=0$, when $f = p/q$ is a rational fraction.  The existence of fractionally charged Abelian anyonic quasiparticles is experimentally established with the observation of  fractional charge combined with an unambiguous theoretical understanding~\cite{dasSarma1996,nayak2008}.   In addition, there is tantalizing and controversial experimental evidence of Abelian and non-Abelian statistics~\cite{nayak2008}.  However, the observation of fractional braiding statistics and the definitive observation of  non-Abelian anyon quasiparticles~\cite{Moore1991} remain elusive--the experimental confirmation of either would be a major step towards the  construction of a topologically protected quantum computing device~\cite{dassarma2005,nayak2008}.

The FQHE requires a quasi-two-dimensional electron system and was first discovered in GaAs semiconductor heterostructures and has since been observed in other quasi-two-dimensional systems, one of which is the  newly discovered atomically thin two-dimensional system of graphene~\cite{Novoselov2004}.   The experimental exploration of the FQHE in graphene  is still in its relatively early development~\cite{Bolotin2009,du2009,dean2011,feldman2012,feldman2013}. Graphene is a hexagonal crystal system of carbon atoms with two atoms ($A$ and $B$ sites) per unit cell. The low-energy Hamiltonian, in the continuum limit of a nearest neighbor tight binding model, consists of $\pi$-electrons in two bands ($K$ and $K^\prime$ valleys) each with a massless linear spectrum, therefore, each two-dimensional electron has a spin and valley index. In the presence of a perpendicular magnetic field, the linear Dirac spectrum gives a cyclotron energy of $\mathrm{sgn}(n)\sqrt{2|n|}\hbar v_F/l_B=(2.2/\epsilon)\;\mathrm{K}$  where $\epsilon$ is the dielectric and  $v_F\sim 10^6$ m/s is the Fermi velocity.  The Landau level index $n = 0,\pm1,\pm 2, \ldots$ has a spacing between consecutive Landau levels decreasing as $1/\sqrt{n}$ for large $n$ (compared to $\hbar\omega_c(n + 1/2)$ for electrons in semiconductor heterostructures with $n = 0, 1, 2, \ldots$ with constant Landau level spacing). 

At the simplest level, one can theoretically study the FQHE with a Hamiltonian consisting of only the Coulomb interaction between electrons in the $n$th Landau level.  However, it is important to take into account realistic physics when they may produce qualitatively different effects compared to the minimal model of the Coulomb Hamiltonian alone.  To leading order, the most important realistic effects in graphene are Landau level mixing and disorder.  (Note that graphene is atomically thin, so unlike the FQHE in semiconductor heterostructure, one does not need to consider the width of the quasi-two-dimensional system.)  Landau level mixing is the tendency of electron/hole excitations in unoccupied/occupied Landau levels outside the $n$th level and can be parameterized by the ratio $\kappa$ of the Coulomb interaction strength to the Landau level spacing: 
\begin{eqnarray}
\kappa = \frac{(e^2/\epsilon l_B)}{(\hbar v_F/l_B)}=e^2/\epsilon \hbar v_F \nonumber
\end{eqnarray}
 and, interestingly, it is independent of the magnetic field strength.  If $\kappa\ll 1$ then Landau level mixing can be safely ignored when constructing an effective theoretical model. Experimental samples where the FQHE in graphene has been observed (both suspended graphene and graphene on a boron nitride substrate~\cite{Bolotin2009,du2009,dean2011,feldman2012,feldman2013}), however, have a Landau level mixing parameter of $0.5 \lesssim \kappa \leq 2.2$ and Landau level mixing can never be safely ignored in graphene.  Therefore, it is important to at least study a well-defined model where the effects of Landau level mixing can be understood in a controlled approximation that is exact in some limit (in our case  as $\kappa\rightarrow 0$).

Previous numerical work~\cite{Apalkov2006,Goerbig2006,Toke2006,Nomura2006,Papic2009,Sodemann2014,Wu2014,Balram2015} has shown the system to be  sensitive to small perturbations to the Hamiltonian and only some~\cite{Peterson2014a} have attempted to take Landau level mixing into account.  In this work, however, we do not discuss specific results of exact diagonalization or variational Monte Carlo studies of the FQHE in graphene, rather, we seek to provide a more accurate formalism going forward in which to investigate realistic effects with less chance of significant systematic errors.

A technique commonly used in theoretical studies is to map the two-dimensional plane to the compact sphere--this geometry has the advantage of being free of boundaries allowing a more straightforward study of bulk properties (we will discuss the spherical geometry in more detail below).  Most numerical studies of the FQHE in graphene that have utilized the spherical geometry have formulated the  Hamiltonian describing the electron-electron interactions in terms of Haldane pseudopotentials calculated in the infinite planar geometry. While it is feasible that the use of planar pseudopotentials in spherical geometry calculations may better approximate the thermodynamic limit, when the energy difference between competing FQH states is small, which is apparently the case for the FQHE in graphene, it is important to carefully approach the thermodynamic limit using spherical geometry pseudopotentials.  
Recent works by Balram \textit{et al}.~\cite{Balram2015} and W{\'o}js \textit{et al}.~\cite{Wojs2011}  have investigated graphene using the spherical pseudopotentials and have provided a formula for the Coulomb matrix elements in the spherical geometry in terms of the usual matrix elements for massive electrons--this allows one to calculate the graphene spherical pseudopotentials. While the mathematical physics problem of Dirac fermions in the presence of a magnetic monopole has received  attention (cf. Refs.~\onlinecite{Kazama1977,Castillo1997,Schliemann2008} and~\onlinecite{Newman1962,Dray1985}) the recent work~\cite{Balram2015,Wojs2011} was justified by appealing to a calculation of the eigenstates by Jellal~\cite{Jellal2008}.

In this work, we accomplish essentially three things: 

(i) One thing we do to provide an alternative derivation (compared to Jellal~\cite{Jellal2008}) of the eigenfunctions and eigenenergies for massless Dirac fermions on the Haldane sphere -- our approach is more in line with the traditional approach used in the FQHE literature and utilizes the cyclotron motion operators discussed previously by Greiter~\cite{Greiter2011}.   Incidentally, we note that our Hamiltonian is different from that analyzed previously~\cite{Schliemann2008}.  The single-electron eigenfunctions $\Psi_{Qnm}$ and eigenenergies $E_{Qn}$ are
\begin{eqnarray}
\Psi_{Qnm}&=&\frac{(\sqrt{2})^{\delta_{n0}}}{\sqrt{2}}
\left( \begin{array}{c}
-\mathrm{sgn}(n)i\mathcal{Y}_{|Q|+1|n|-1m} \\
\mathcal{Y}_{|Q||n|m} \end{array} \right)\nonumber
\end{eqnarray}
and
\begin{eqnarray}
E_{Qn} &=& \mathrm{sgn}(n)\frac{\hbar v_F}{l_B}\sqrt{2|n|+\frac{|n|(|n|+1)}{Q}}\nonumber
\end{eqnarray}
where $\mathcal{Y}_{Qnm}$ are the monopole harmonics used in FQHE studies in the spherical geometry, $Q$ is the monopole strength at the center of the sphere that produces the radial magnetic field, and $n=0,1,2,\ldots$ is the Landau level index.    

(ii) The second thing we do is to use the  above single-particle eigenstates $\Psi_{Qnm}$ to calculate the Haldane pseudopotentials for the $n=1$ Landau level of graphene completely within the spherical geometry and tabulate the values for a number of commonly diagonalized or studied system sizes.   

(iii) The third thing we do is formulate an effective Landau level mixing Hamiltonian entirely within the spherical geometry for use in subsequent studies. This is possible because  we find the single-particle kinetic energy ($E_{Qn}$).  This is a crucial ingredient to understand the effect of Landau level mixing in graphene for finite-sized spherical systems.   Again, we tabulate the three-body pseudopotentials and two-body pseudopotential corrections that characterize the effective interaction for a number of commonly studied finite-sized systems.  

In the process of characterizing the finite-sized effective Hamiltonian, we learn a number of important things.  We learn precisely how the pseudopotentials approach the thermodynamic limit, how different the finite-size values are from the values in the thermodynamic limit, and the number of Landau levels that need to be kept in the sums over virtual transitions to higher and lower Landau levels in order to obtain proper convergence.  This is important because an alternative approach to studying Landau level mixing in the FQHE is to exactly diagonalize or implement density-matrix-renormalization-group techniques in an expanded, yet  truncated, Fock space~\cite{ZaletelPRB2015}.  However, due to computational limitations, the number of Landau levels kept in these sorts of calculations is on the order of five or six.  What we learn here is that the three-body pseudopotentials converge rather quickly with the number of Landau levels kept in the sums and usually are nearly converged by five or six Landau levels--this is good news for the expanded Fock space method of incorporating Landau level mixing.  However, the two-body corrections to the bare pseudopotentials commonly require well over ten Landau levels to ensure reasonable convergence--this is not very good news for the expanded Fock space approach.  It is important in the future to determine the validity and precise parameter regimes where the two alternative methods of including Landau level mixing are valid.  

This paper is organized  as follows:  in Sec.~\ref{diracsphere} we derive the eigenvalues and eigenfunctions for massless Dirac fermions on the Haldane sphere, in Sec.~\ref{spherevms} we analyze the Haldane pseudopotentials in the $n=1$ Landau level, in Sec.~\ref{llmixing} we formulate the effective Landau level mixing Hamiltonian for graphene entirely within the spherical geometry and provide the two-body pseudopotential corrections and three-body pseudopotentials, 
in Sec.~\ref{diag} we compare results of exact diagonalization by using the newly derived spherical pseudopotentials against results using the pseudopotentials calculated in the infinite planar geometry, and, finally in Sec.~\ref{conc} we conclude.  For completeness we provide some derivations and formulas in appendixes~\ref{app:matrixelement} and~\ref{app:map}.

\section{Landau levels for massless Dirac fermions in the spherical geometry}
\label{diracsphere}

\begin{figure}[t]
\begin{center}
\includegraphics[width=7.5cm]{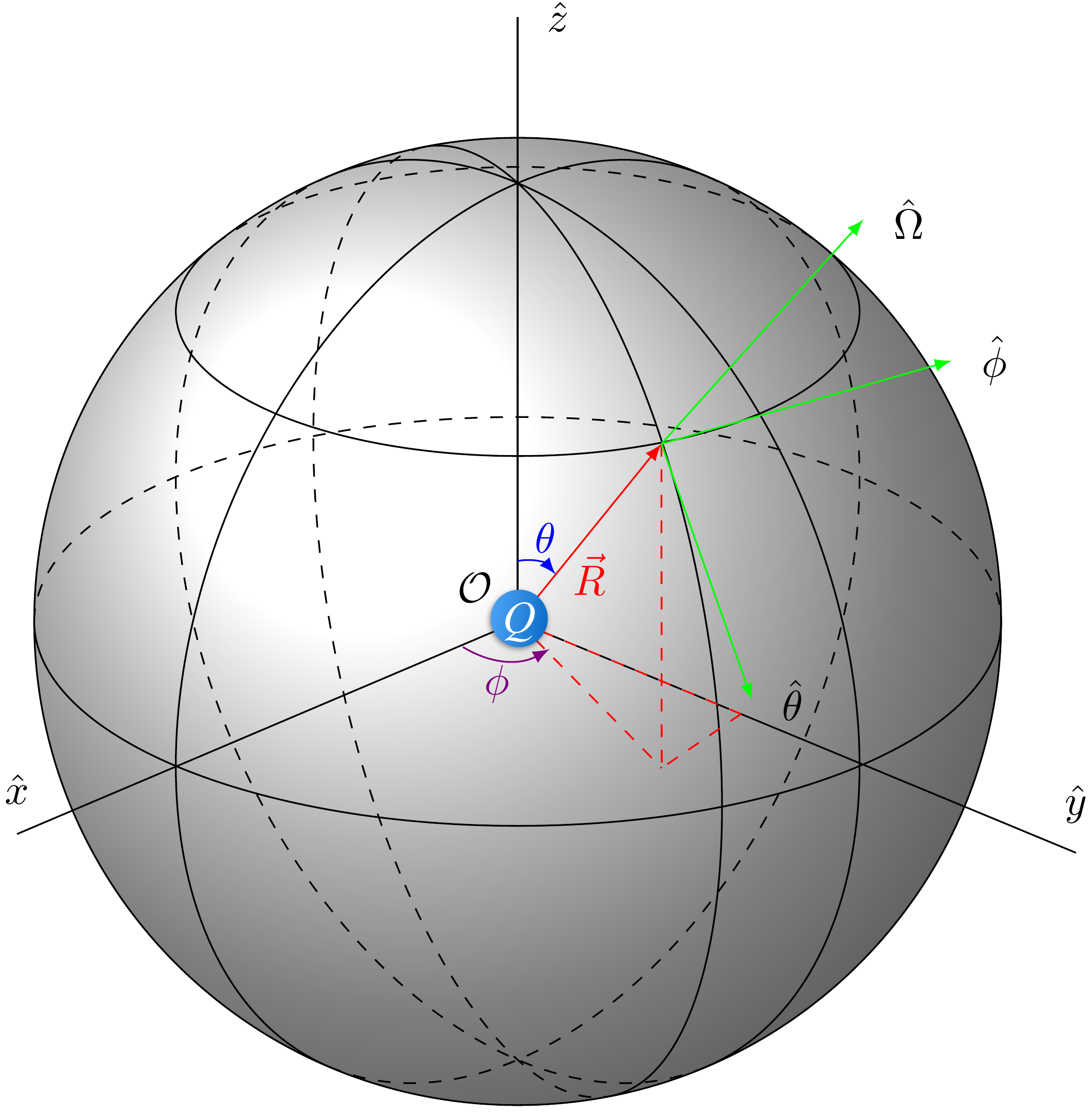}
\caption{The Haldane sphere:  a magnetic monopole of strength $Q$ is placed at the center of a sphere of radius $R=\sqrt{Q}l_B$ producing a radial magnetic field of strength $B=\hbar c Q/eR^2$.  The radial unit vector $\hat\Omega=\mathbf{R}/R$ is shown in green in addition to the coordinates in the tangent plane (the plane defined by $\hat\theta$ and $\hat\phi$.)  
 }  
\label{fig:sphere}
\end{center}
\end{figure}

We wish to calculate the single particle eigenvalues and eigenfunctions for massless Dirac fermions confined to the surface of a sphere of radius $R$ in the presence of a magnetic monopole of strength $Q=R^2/\l_B^2$, i.e., we confine the particles to the so-called Haldane sphere~\cite{haldane1983}.  We choose the vector potential $\mathbf{A}=-\hat\phi Q c \cot(\theta)/eR$ such that $\nabla\times\mathbf{A}=B\hat\Omega$ where $\hat\Omega=\mathbf{R}/R$ is the unit vector in the radial direction.  The single particle solution for massive fermions with a quadratic energy dispersion are known and the eigenfunctions are given by the monopole harmonics $Y_{Qlm}(\theta,\phi)$ where $m=-l,-l+1,\ldots,l-1,l$ is the $z$-component of angular momentum, $l=|Q|+|n|$ is the single particle  angular momentum,  the Landau level (LL) index $n=0,1,2,\ldots$, and $\theta$ and $\phi$ are the polar and azimuthal angles, respectively~\cite{WuYang1976,haldane1983,jainCF}.  See Fig.~\ref{fig:sphere} for an illustration.

\subsection{Review of solution for massless fermions on the plane}
We briefly review the solution of the Landau problem in the planar geometry, which has been shown  before~\cite{Apalkov2006,Nomura2006,Toke2006,Goerbig2006,Toke2007}, to ease the discussion of the spherical geometry solution that follows.  The low-energy Hamiltonian for electrons in graphene is 
\begin{eqnarray}
\label{hgraph}
H=v_F
\left( \begin{array}{cc}
0 & \Pi_x - i\Pi_y \\
\Pi_x + i\Pi_y & 0 \end{array} \right)
= v_F\mathbf\sigma\cdot\mathbf\Pi
\end{eqnarray}
where $\mathbf\sigma=(\sigma_1,\sigma_2,\sigma_3)$ are the Pauli matrices, and $\mathbf\Pi=\mathbf{p}+(e/c)\mathbf{A}$ is the canonical momentum with $\mathbf{A}$ being the vector potential satisfying $\nabla\times\mathbf{A}=B\hat{z}$.  
After introducing ladder operators $a^\dagger=(il_B/\hbar\sqrt{2})(\Pi_x+i\Pi_y)$ and $a=-(il_B/\hbar\sqrt{2})(\Pi_x-i\Pi_y)$, such that $[a,a^\dagger]=1$, we can rewrite the Hamiltonian as
\begin{eqnarray}
H=\frac{\sqrt{2}\hbar v_F}{il_B}
\left( \begin{array}{cc}
0 & a \\
-a^\dagger & 0 \end{array} \right)\;.
\end{eqnarray}
Amusingly, the square of $H$ is diagonal, i.e., 
\begin{eqnarray}
H^2=\frac{2\hbar^2 v_F^2}{l^2_B}
\left( \begin{array}{cc}
aa^\dagger & 0 \\
0 & a^\dagger a \end{array} \right)
\end{eqnarray}
and the eigenfunctions  of $H^2$ can be readily found to be
\begin{eqnarray}
\label{eq:eigs-plane}
\psi_{nm}(x,y)=\frac{(\sqrt{2})^{\delta_{n0}}}{\sqrt{2}}
\left( \begin{array}{c}
-\mathrm{sgn}(n)i\eta_{|n|-1,m}(z) \\
\eta_{|n|m}(z) \end{array} \right)
\end{eqnarray}
where $\eta_{nm}(z)$ are the single-particle eigenfunctions of the usual quadratic energy dispersion for massive fermions (for example, electrons in a GaAs heterostructure) with $n=0,1,2,\ldots$ the LL index, $m=-n,-n+1,\ldots,0,1,\ldots$ the orbital angular momentum (cf. Ref.~\onlinecite{jainCF}),  $\mathrm{sgn}(0)=0$,  $\mathrm{sgn}(x)=-1$ for $x<0$, and $\mathrm{sgn}(x)=1$ for $x>0$.  Since $[H,H^2]=0$ the eigenfunctions of $H$ are given by $\psi_{nm}(x,y)$ [Eq.(~\ref{eq:eigs-plane})] and the eigenenergy is 
\begin{eqnarray}
E_{n}=\hbar v_F\sqrt{2|n|}/l_B\;.
\end{eqnarray}

\subsection{Review of Landau problem for massive fermions on the sphere}
We now  review the solution for massive fermions with quadratic dispersion confined to the surface of the Haldane sphere.We take Greiter's lead and introduce cyclotron motion operators $\mathbf{S}=(S_1,S_2,S_3)$~\cite{Greiter2011}--these operators are essentially the operators for rotations in terms of Euler angles in the body-fixed frame compared to the usual angular-momentum operators which are in terms of Euler angles in the space-fixed frame.  These are most easily formulated by using Haldane's spinor coordinates $u\equiv\cos(\theta/2)\exp(i\phi/2)$ and $v\equiv\sin(\theta/2)\exp(-i\phi/2)$ as
\begin{eqnarray}
S_-&=&S_1+iS_2=\hbar\left(u\frac{\partial}{\partial\bar{v}}-v\frac{\partial}{\partial\bar{u}}\right)\\
S_+&=& S_1-iS_2=\hbar\left(\bar{v}\frac{\partial}{\partial u}-\bar{u}\frac{\partial}{\partial v}\right)\\
S_3&=&\frac{\hbar}{2}\left(u\frac{\partial}{\partial u}+v\frac{\partial}{\partial v}-\bar{u}\frac{\partial}{\partial\bar{u}}-\bar{v}\frac{\partial}{\partial\bar{v}}\right)\;.
\end{eqnarray}
The cyclotron operators obey the algebra $[S_i,S_j]=i\hbar\epsilon_{ijk}S_k$ and we further note that $[S_i,L_j]=0$ for all $i$ and $j$ where the $L_x$, $L_y$, and $L_z$ are the components of the angular-momentum operator $\mathbf{L}$ .  All the operators $H,\;S^2,\;L^2,\;S_3,\;L_3$  mutually commute and share  common eigenfunctions which are the monopole harmonics $Y_{Qlm}(\theta,\phi)$ mentioned above.  Since the $Y_{Qlm}$'s are eigenfunctions of $\mathbf{S}$, we can calculate their eigenvalues.  First we change the notation of the monopole harmonics and write the $Y_{Qlm}$ in such a way to more easily facilitate our final answer in the graphene case.   Let us define
\begin{eqnarray}
\label{y-defn}
\mathcal{Y}_{Qnm} \equiv Y_{Q,Q+n,m}=Y_{Qlm}
\end{eqnarray}
to more clearly display the LL index quantum number $n$.  The above operators  $\mathbf{S}$  act on the $\mathcal{Y}_{Qnm}$ in the following ways:
\begin{eqnarray}
S^2 \mathcal{Y}_{Qnm}&=&\hbar^2(Q+n)(Q+n+1)\mathcal{Y}_{Qnm}\;,\\
S_3 \mathcal{Y}_{Qnm}&=&\hbar Q\mathcal{Y}_{Qnm}\;,\\
S_\pm \mathcal{Y}_{Qnm}&=&\hbar \sqrt{(Q+n)(Q+n+1)-Q(Q\pm 1)}\nonumber\\
&&\;\;\;\times\mathcal{Y}_{Q\pm1n\mp1m}\;.
\end{eqnarray}
We see that $S_\pm$ lowers (raises) the LL index $n$ while simultaneously raising (lowering) the monopole strength $Q$.  The single-particle angular momentum $l=Q+n$ remains constant throughout all the above operations.  

For massive fermions the single-particle Hamiltonian is 
\begin{eqnarray}
H=\frac{\mathbf\Pi^2}{2m}\;.
\end{eqnarray}
For fermions confined to the surface of a sphere of radius $R=\sqrt{Q}l_B$, the two-dimensional ``plane" is the plane tangent to the spherical surface.  We can define the components of the canonical momentum tangent to the plane through  $\mathbf\Pi = \hat\Omega\times[i\hbar\nabla + (e/c)\mathbf{A}]$.  By using the definition of $\mathbf{A}$ above,
\begin{eqnarray}
\Pi_\theta &=& -i\hbar\frac{1}{R\sin(\theta)}\frac{\partial}{\partial\phi} + \frac{1}{R} \cot(\theta)S_3\\
\Pi_\phi &=& -i\hbar\frac{1}{R}\frac{\partial}{\partial \theta}\;.
\end{eqnarray} 
Note that above we replaced $Q$ in $\mathbf{A}$ with the operator $S_3$ because $Q$ is its eigenvalue.  

It turns out that, after some algebra (see Appendix~\ref{app:map}), we can relate $\mathbf{S}$ and $\mathbf\Pi$ through $R\Pi_\theta = -S_1$ and $R\Pi_\phi=S_2$.  This formulation is more natural if we are thinking of fermions confined to the surface of a sphere with a radial magnetic field as the map of the planar system to the spherical one--compared to some combination of $\Pi_x$, $\Pi_y$, and $\Pi_z$ or in terms of $\mathbf{L}$.  Now we can write
 \begin{eqnarray}
 H&=&\frac{\mathbf\Pi^2}{2m}= \frac{\Pi_\theta^2 + \Pi_\phi^2}{2m}\nonumber\\
 &=&\frac{(\Pi_\theta + i\Pi_\phi)(\Pi_\theta - i\Pi_\phi) - i[\Pi_\phi,\Pi_\theta]}{2m}\nonumber\\
 &=&\frac{S_-S_+-i[S_2,-S_1]}{2mR^2}\nonumber\\
 &=&\frac{S_-S_+  + \hbar S_3}{2mR^2}\\
 &=& \frac{\omega_c}{\hbar}\left(\frac{S_-S_+}{2Q} + \hbar\frac{S_3}{2Q}\right)
 \label{hmasssphere}
 \end{eqnarray} 
where have substituted $R^2=Q l_B^2$ in the last line and introduced the cyclotron frequency $\omega_c=eB/mc$.  Remembering that the eigenvalue of $S_3$ is $Q$ we see that this is in direct analogy to the planar system where $H=\hbar\omega_c (a^\dagger a + 1/2)$ because $S_-S_+$ is basically the number operator in the spherical geometry.  The action of $S_-S_+$ on $\mathcal{Y}_{Qnm}$ is 
 \begin{eqnarray}
S_-S_+\mathcal{Y}_{Qnm}  &=&[n(n+1)+2nQ]\hbar^2\mathcal{Y}_{Qnm}\;.
 \end{eqnarray}
 Hence, the eigenvalue of Eq.~(\ref{hmasssphere}) is the well-known result
 \begin{eqnarray}
 E_{Qn}=\hbar\omega_c\left(n+\frac{1}{2} +\frac{ n(n+1)}{2Q}\right)\;.
 \end{eqnarray}
 In the thermodynamic limit we obtain the planar result; $E_n=\lim_{Q\rightarrow\infty}E_{Qn}=\hbar\omega_c(n+1/2)$.

\subsection{Solution for massless fermions on the sphere}
We now tackle the graphene problem.  From Eq.~(\ref{hgraph}) we write, expanding the Pauli matrices,
\begin{eqnarray}
H&=&v_F\sigma\cdot\mathbf{\Pi}\nonumber\\
&=&v_F \left( \begin{array}{cc}
0 & \Pi_\theta - i \Pi_\phi \\
\Pi_\theta + i \Pi_\phi & 0 \end{array} \right)\;.
\end{eqnarray}
This formulation of the Hamiltonian is the most natural for graphene on the Haldane sphere because the dynamical momentum of the electrons is  tangent to the spherical surface (in the tangent plane).  Equipped with the cyclotron operators $S_-$, $S_+$, and $S_3$ we can now simply follow the procedure used in the planar geometry to readily obtain the eigenfunctions and eigenvalues.  The Hamiltonian is
\begin{eqnarray}
\label{hgraphsphere}
H&=&-\frac{v_F}{R}\left( \begin{array}{cc}
0 & S_1 + i S_2 \\
S_1 - i S_2 & 0 \end{array} \right)\nonumber\\
&=&-\frac{v_F}{R}\left( \begin{array}{cc}
0 & S_+ \\
S_- & 0 \end{array} \right)\;.
\end{eqnarray}
Again, the square of $H$ is diagonal
\begin{eqnarray}
\label{hsquaregraphsphere}
H^2&=&\frac{ v_F^2}{R^2}
\left( \begin{array}{cc}
S_+S_- & 0 \\
0 & S_-S_+ \end{array} \right)\nonumber\\
&=&\frac{ v_F^2}{R^2}
\left( \begin{array}{cc}
S_-S_++2\hbar S_3 & 0 \\
0 & S_-S_+ \end{array} \right)\;.
\end{eqnarray}
Hence, in direct analogy to the planar system, we can find the eigenfunctions  of Eq.~(\ref{hsquaregraphsphere} ) [and hence Eq.~(\ref{hgraphsphere})]
\begin{eqnarray}
\label{eq:eigs-sphere}
\Psi_{Qnm}=\frac{(\sqrt{2})^{\delta_{n0}}}{\sqrt{2}}
\left( \begin{array}{c}
-\mathrm{sgn}(n)i\mathcal{Y}_{|Q|+1|n|-1m} \\
\mathcal{Y}_{|Q||n|m} \end{array} \right)\;,
\label{efns}
\end{eqnarray}
and the eigenvalues of Eq.~(\ref{hgraphsphere}) are 
\begin{eqnarray}
\label{eq:energraphsphere}
E_{Qn} = \mathrm{sgn}(n)\frac{\hbar v_F}{l_B}\sqrt{2|n|+\frac{|n|(|n|+1)}{Q}}\;.
\end{eqnarray}
In the thermodynamic limit the planar result is obtained, $\lim_{Q\rightarrow\infty}E_{Qn}=\mathrm{sgn}(|n|)\hbar v_F\sqrt{2|n|}/l_B$.

An interesting feature of the graphene eigenfunctions on the plane is that for $n\neq0$ the electron is partially in the $n$th LL and the $(n-1)$st LL.  In the spherical geometry this is also true but the single particle angular momentum $l=|Q|+|n|$ is a good quantum number and constant for both electron components--the value of the monopole harmonic is shifted by one unit to compensate.  That is, for the component in the $(n-1)$st LL the monopole strength is $|Q|+1$ while for the component in the $n$th LL the monopole strength remains $Q$. 
This has led some~\cite{Balram2015} to define an average magnetic length through $l^\mathrm{av}=R/\sqrt{Q^\mathrm{av}}$, where $Q^\mathrm{av}$ is the average flux, since the spherical radius is related to the square root of the monopole strength.  However, in our treatment the magnetic length is well defined through $R=l_B\sqrt{Q}$ with no ambiguity.

\section{Bare Haldane Pseudopotentials}
\label{spherevms}

\begin{figure}[t]
\begin{center}
\includegraphics[width=9.cm]{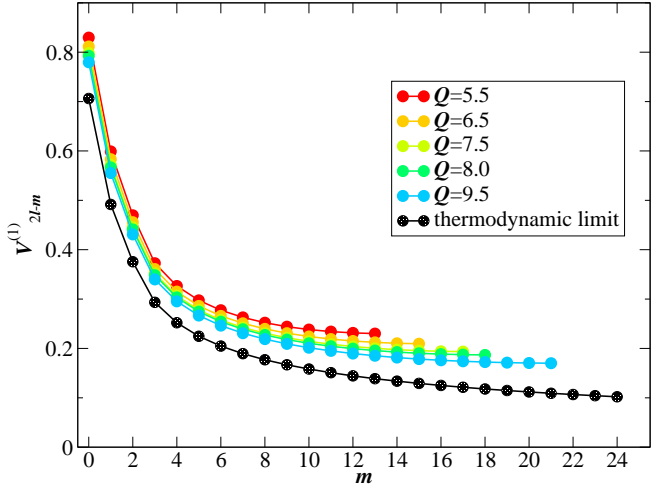}
\caption{Spherical pseudopotentials for the $n=1$ Landau level of graphene.  Recall that the $n=0$ graphene pseudopotentials are identical to those for massive fermions.  We plot $V^{(1)}_{2l-m}$  vs $m$ for a few notable values of $2l$ (or $Q$ where $l=Q+1$) in addition to the planar value (thermodynamic limit).  See Table~\ref{table1} for specific values.  
 }  
\label{fig:n1vms}
\end{center}
\end{figure}

The  many-body Hamiltonian for interacting massless Dirac fermions  on the sphere is given by the Coulomb interaction and parametrized by the Haldane pseudopotentials
\begin{eqnarray}
H&=&\sum_{i<j}V(\mathbf{r}_i,\mathbf{r}_j)=\sum_{i<j}\frac{e^2}{\epsilon |\mathbf{r}_i-\mathbf{r}_j|}\nonumber\\
&=&\sum_{m=0}^{2l}V^{(n)}_{2l-m}\sum_{i<j}P_{ij}(2l-m)
\end{eqnarray}
where $P_{ij}(2l-m)$ is a projection operator that projects onto states with relative angular momentum $2l-m$ and $V^{(n)}_{2l-m}$ are the Haldane pseudopotentials, i.e., the Coulomb energy between two electrons with relative angular momentum $2l-m$;  note that relative angular momentum $m$ in the planar geometry maps to $2l-m$ in the spherical geometry, i.e., $\lim_{l\rightarrow\infty}V^{(n)}_{2l-m}=\bar{V}^{(n)}_m$ where $\bar{V}^{(n)}_m$ are the pseudopotentials in the infinite plane.  It is common to take the distance between two electrons on the sphere to be the chord distance equal to $|\mathbf{r}_i-\mathbf{r}_j|=\sqrt{2}R|u_1v_2-u_2v_1|$.  
By using the single-particle eigenfunctions for massless Dirac fermions above [Eq.~(\ref{efns})] we can explicitly write 
\begin{widetext}
\begin{eqnarray}
\label{bare}
V^{(n)}_{2l-m} &=& \sum_{\{m_i\}} \langle l,m_1;l,m_2 | 2l-m,m_1+m_2\rangle \langle l,m_3;l,m_4 | 2l-m,m_3+m_4\rangle \nonumber\\
&&\times\delta_{m_1+m_2,m_3+m_4}\delta_{m_1+m_2,2l-m}\langle nm_4,nm_3|V|nm_2,nm_1\rangle^{(Q,n)}_\mathrm{graph}
\end{eqnarray}
where
\begin{eqnarray}
\langle n_4m_4,n_3m_3|V|n_2m_2,n_1m_1\rangle^{(Q,n)}_\mathrm{graph} &=& \frac{(\sqrt{2})^{\sum_{i=1}^4\delta_{n_i 0}}}{4}
\big(\langle |n_4|m_4,|n_3|m_3|V||n_2|m_2,|n_1|m_1\rangle^{(Q,n)}\nonumber\\
&+& \mathrm{sgn}(n_4 n_2)\langle |n_4|-1m_4,|n_3|m_3|V||n_2|-1m_2,|n_1|m_1\rangle^{(Q,n)}  \nonumber\\
&+&\mathrm{sgn}(n_3 n_1)\times\langle |n_4|m_4,|n_3|-1m_3|V||n_2|m_2,|n_1|-1m_1\rangle^{(Q,n)} \nonumber\\
&+&\mathrm{sgn}(n_4 n_3 n_2 n_1)\times\langle |n_4|-1m_4,|n_3|-1m_3|V||n_2|-1m_2,|n_1|-1m_1\rangle^{(Q,n)} \big)\;.\nonumber\\
\label{eq:matrixelement}
\end{eqnarray}
\end{widetext}
$V = V(\mathbf{r}_1,\mathbf{r}_2)$, and $\langle n_4m_4,n_3m_3|V|n_2m_2,n_1m_1\rangle^{(Q,n)}$ is the general two-body Coulomb interaction matrix element given for completeness in Appendix~\ref{app:matrixelement}.  Note that the superscript $(Q,n)$ is to indicate that this matrix element is taken between states of constant angular momentum $l=Q+n$ and each of the sums over the $m_i$'s in $\sum_{\{m_i\}}$ go from $-l$ to $l$.  In Table~\ref{table1} we give the values of  $V^{(1)}_{2l-m}$ for the $n=1$ LL for a number of system sizes of interest and in Fig.~\ref{fig:n1vms} we plot them versus $m$.  In particular we provide pseudopotentials for a few commonly studied systems, i.e.,  $2l=13, 15, 17, 18$, and $21$.  These system sizes can be used to study the Moore-Read Pfaffian~\cite{Moore1991} (for a 1/2 filled LL) and Laughlin~\cite{laughlin1983} (for a 1/3 filled LL) states projected into the $n=1$ LL for $N=8$, 10, 12 and $N=6$, 7, 8 electrons, respectively.  (The relationship between the total flux $2l$ on the sphere and the particle number $N$ for the Moore-Read Pfaffian and Laughlin states is $2l=2N-3$ and $2l=3(N-1)$, respectively).  
Last, we note that we do not provide any pseudopotentials for the lowest $n=0$ LL since they are identical to those for massive fermions given elsewhere.

\begin{table}
\caption{$V_m$ for a few common values for $2l=2|Q|+2|n|$'s (see text).  Below we take $n=1$ for all values.  Hence, the monopole strength $Q=2l/2-1$ and not, simply, $2l/2$.  All energies are given in units of $e^2/\epsilon l_B$. The pseudopotentials in the thermodynamic limit (planar geometry) are $\overline{V}^{(1)}_m$.}
\label{table1}
\begin{tabular}{c|ccccc|l}
\hline\hline
$m$ & $V^{(1)}_{13-m}$ & $V^{(1)}_{15-m}$ & $V^{(1)}_{17-m}$ & $V^{(1)}_{18-m}$ &  $V^{(1)}_{21-m}$ &  $\overline{V}^{(1)}_{m}$ \\
\hline 
0 & 0.829596 & 0.811619 & 0.798223 & 0.792728 &  0.779586   &  0.706212 \\
1 & 0.599088 & 0.583274 & 0.571519 & 0.566705 &   0.555210   &  0.491579 \\
2 & 0.469699 & 0.455676 & 0.445291 & 0.441048 &   0.430940  &  0.375608 \\
3 & 0.372646 & 0.360723 & 0.351917 & 0.348325 &   0.339783  &  0.293390 \\
4 & 0.326480 & 0.315110 & 0.306749 & 0.303347 &   0.295277  &  0.251956 \\
5 & 0.297444 & 0.286145 & 0.277880 & 0.274528 &   0.266602   &  0.224640 \\
6 & 0.277387 & 0.265888 & 0.257531 & 0.254154 &    0.246199 &  0.204748 \\
7 & 0.262864 & 0.250969 & 0.242390 & 0.238939 &   0.230841  &  0.189393 \\
8 & 0.252102 & 0.239646 & 0.230741 & 0.227176 &   0.218852  &  0.177064 \\
9 & 0.244088 & 0.230914 & 0.221589 & 0.217877 &  0.209256   &  0.166877 \\
10 & 0.238205 & 0.224149 & 0.214313 & 0.210423 &   0.201440  & 0.158273 \\
11 & 0.234064 & 0.218948 & 0.208506 & 0.204406 &   0.194999  & 0.150880 \\
12 & 0.231421 & 0.215039 & 0.203889 & 0.199546 &  0.189652   & 0.144437 \\
13 & 0.230133 & 0.212242 & 0.200268 & 0.195646 &   0.185200  & 0.138755 \\
14 & 		         & 0.210438 & 0.197507 & 0.192566 &   0.181496  & 0.133697 \\
15 & 		         & 0.209553 & 0.195510 & 0.190204 &  0.178431   &  0.129155 \\
16 & 		         &		   & 0.194212 & 0.188490 &  0.175925   &  0.125047 \\
17 & 		         &		   & 0.193573 & 0.187372 &   0.173915  & 0.121308 \\
18 &  	         & 		   &  		      & 0.186820 &   0.172356  &  0.117886 \\
19 &  		& 		   &  		      &  		&  0.171214   &  0.114738 \\
20 &  		&		   &  		      &  		&  0.170465  & 0.111830 \\
21 &  		&		   &  	               &  		&  0.170095   & 0.109132 \\
22 &  		&		   &  		      & 		&     			& 0.106621 \\
23 &  		&		   &  		      &  		&    			 & 0.104276 \\
24 &  		&		   &  		      &  		&     			& 0.102079 \\
\hline
\hline
\end{tabular}
\end{table}

\section{Landau level mixing: Haldane Pseudopotential Corrections}
\label{llmixing}

Landau level mixing occurs when the electrons that partially fill the $n$th LL  have a
significant probability amplitude of making virtual transitions to higher unoccupied and  lower occupied LLs due to the Coulomb interaction.  We focus on systems where the LLs of spin and valley internal degrees of freedom are approximately degenerate.  As mentioned above the tendency for LL mixing is captured in the LL mixing parameter given by the ratio of the Coulomb interaction energy to the cyclotron energy,
\begin{eqnarray}
\label{kappa}
\kappa =\frac{\left(\frac{e^2}{\epsilon l_B}\right)}{\left(\frac{\hbar v_F}{l_B}\right)}=\frac{e^2}{\epsilon \hbar v_F}=\frac{2.2\;\mathrm{(Kelvin)}}{\epsilon}
\end{eqnarray}
by using $v_F=10^6$ m/s in the last equality.  Since this has no magnetic field $B$ dependency it can only be suppressed through the manipulation of the dielectric $\epsilon$.  For current experimental systems $0.5\lesssim \kappa \leq 2.2$.

It is a difficult theoretical problem to include LL mixing within exact diagonalization.  For graphene it is particularly difficult since, without explicit or spontaneous symmetry breaking of the SU(4) valley and spin degeneracy, the Hilbert space is formidably large.  It is therefore beyond current computational capabilities of exact diagonalization to expand the Hilbert space and allow electrons (holes) in the $n$n$th^\mathrm{th}$ LL to occupy unoccupied (occupied) Landau levels outside this level.  

To approximately include LL mixing, one of the current authors (along with Nayak) obtained a realistic effective Hamiltonian taking into account LL mixing perturbatively in powers of the LL  mixing parameter $\kappa$ following the original work of Ref.~\onlinecite{Bishara2009}.  An advantage of our approach, outlined in Ref.~\onlinecite{Peterson2013}, is that it is exact in the $\kappa\rightarrow 0$ limit.  The disadvantage, or course, is that it is perturbative and our small parameter $\kappa$ is not necessarily always small [cf. Eq.~(\ref{kappa})].  Ultimately we write an effective many-body Hamiltonian in terms of Haldane pseudopotentials
\begin{eqnarray}
\label{eq:LLMham}
H_\mathrm{eff}(\kappa) &=& \sum_{i<j}V_\mathrm{2body}(\kappa,\mathbf{r}_i,\mathbf{r}_j) + \sum_{i<j<k}V_\mathrm{3body}(\kappa,\mathbf{r}_i,\mathbf{r}_j,\mathbf{r}_k)\nonumber\\
&=&\sum_{m=0}^{2l} V^{(n)}_{2l-m,\mathrm{2body}}(\kappa)\sum_{i<j}P_{ij}(2l-m)\nonumber\\
&&+\sum_{m=0}^{3l} V^{(n)}_{3l-m,\mathrm{3body}}(\kappa)\sum_{i<j<k}P_{ijk}(3l-m)
\end{eqnarray}
where $P_{ijk}(3l-m)$ is a projection operator that projects onto triplets of electrons with relative angular momentum $3l-m$.  $V^{(n)}_{2l-m,\mathrm{2body}}(\kappa)$ and $V^{(n)}_{3l-m,\mathrm{3body}}(\kappa)$ are the two- and three-body, $\kappa$ dependent, Haldane pseudopotentials.  The two-body pseudopotential can be written as
\begin{eqnarray}
V^{(n)}_{2l-m,\mathrm{2body}}(\kappa) = V^{(n)}_{2l-m} + \kappa \delta V^{(n)}_{2l-m}
\end{eqnarray}
which is a sum of the (bare) $\kappa$-independent Coulomb pseudopotential [cf. Eq.~(\ref{bare})] plus $\kappa$ times a correction $\delta V^{(n)}_{2l-m}$ due to LL mixing.  In general, LL mixing does two things.  One is that it ``softens" the two-body interactions (in the thermodynamic limit), i.e., $\delta \overline{V}^{(n)}_{m,\mathrm{2body}} < 0$ where $\delta \overline{V}^{(n)}_{m,\mathrm{2body}}$ is the pseudopotential correction in the thermodynamic limit--see below that this is not true for finite-sized spherical systems.  The second thing LL mixing does is generate particle-hole symmetry breaking three-body terms.  

Early theoretical work on the FQHE in graphene, much of it before any experimental observation~\cite{Bolotin2009,du2009,dean2011,feldman2012,feldman2013}, did not consider Landau level mixing and their connection to experiments is therefore tenuous. However, more recent work has unearthed an energy landscape of a variety of possible ground states that are very close in energy~\cite{Peterson2014a,Balram2015}.  Hence it is important for all finite size studies to approach the thermodynamic limit as delicately as possible.  Our contention is that one should use pseudopotentials, and LL mixing corrections to the bare pseudopotentials, fully appropriate to the finite-sized spherical system under study.  To that end, we characterize the LL mixing effective Hamiltonian for graphene for the same systems sizes for which we calculated the bare two-body pseudopotentials above, i.e., we calculate the two- and three-body pseudopotentials for graphene in the presence of LL mixing.

The formalism used to calculate $\delta V^{(n)}_{2l-m,\mathrm{2body}}$ and $V^{(n)}_{3l-m,\mathrm{3body}}(\kappa)$ is provided in Ref.~\onlinecite{Peterson2013} and will not be reproduced here.  The main difference between the previous calculations of the LL mixing effective Hamiltonian in the planar geometry and the one presented here for the spherical geometry is the nature of the sums over angular momenta involved in the virtual transitions across  LLs and the use of the spherical geometry finite-size systems kinetic energy appearing in the denominators of the expressions.   Instead of the angular-momentum sums extending from zero to infinity, the sums now go over the possible single-particle angular momenta available on the sphere, i.e.,  from $-l$ to $l$ where $l=|Q|+|n|$.   For example, the three-body pseudopotential can be found through
\begin{eqnarray}
\label{eq:3bodyVm}
V_{L,\mathrm{3body}}^{(n)}&=&\sum_{\{m_i\}} \langle L,M|l, m_4,\gamma';l, m_5,\beta';l, m_6,\alpha'\rangle\nonumber\\
&&\hspace{-0.75cm}\times\langle l, m_1,\alpha;l, m_2,\beta;l, m_3,\gamma |L,M\rangle u^\mathrm{3body}_{654;321}\;,
\end{eqnarray}
where the $\sum$ indicates a sum over all $\{m_i\}\in[-l,l]$ and primed spin variables $(\alpha',\beta',\gamma')$ with
\begin{eqnarray}
\label{eq:u3body}
u^\mathrm{3body}_{654;321}=-\sideset{}{'}\sum_{n_x=-\infty}^\infty \sum_{m_x=-l_x}^{l_x} \sum_{\gamma=\uparrow,\downarrow}\sum_\mathrm{cyc. perm.}\frac{V^{\alpha'\lambda,\beta\alpha}_{6x,21}V^{\beta'\gamma',\lambda\gamma}_{54,x3}}{E_{Qn_x}-\mu}\;,\nonumber\\
\end{eqnarray}
where $l_x=|Q|+|n_x|$,  $\mu=E_{Qn}$ is the chemical potential, and the prime on the sum over $n_x$ indicates that we do not include $n_x=n$.  The energies in the denominator are of course given by our new expression for the spherical kinetic energy [Eq.~(\ref{eq:energraphsphere})].  The matrix elements are
\begin{eqnarray}
V_{43,21}^{\beta'\alpha',\beta\alpha} = V_{43,21}\delta^{\alpha\alpha'}\delta^{\beta\beta'}-V_{34,21}\delta^{\alpha\beta'}\delta^{\beta\alpha'}
\end{eqnarray}
where $\alpha$, $\alpha'$, $\beta$, and $\beta'$ label the spin indices.  The Coulomb matrix element is 
\begin{eqnarray}
V_{43,21}=\langle n_4m_4,n_3m_3|V|n_2m_2,n_1m_1\rangle^Q_\mathrm{graph}
\end{eqnarray}
given in Eq.~(\ref{eq:matrixelement}) and the matrix elements for the spherical geometry are well known~\cite{jainCF} and given in Appendix~\ref{app:matrixelement} for completeness.  We encourage the reader to consult Ref.~\onlinecite{Peterson2013} for more details regarding the formalism for calculating the pseudopotentials characterizing the realistic effective LL mixing Hamiltonian for graphene.  The modifications described above for calculating $V_{L,\mathrm{3body}}^{(n)}$ purely within the spherical geometry are straightforward and easily generalized for the two-body pseudopotential corrections $\delta V^{(n)}_{L,\mathrm{2body}}$.    Finally, we briefly point out that, in the spherical geometry,  the relative angular momentum $L$ maps to a relative angular momentum of $m$ in the planar geometry.  That is, for the two-body and three-body terms we have $L=2l-m$ and $L=3l-m$ mapping to $m$, respectively, i.e., 
\begin{eqnarray}
\lim_{Q\rightarrow \infty}[V^{(n)}_{2(|Q|+|n|)-m,\mathrm{3body}}]_\mathrm{sphere}=[\delta V^{(n)}_{m,\mathrm{2body}}]_\mathrm{plane}\;,\nonumber
\end{eqnarray}
and
\begin{eqnarray}
\lim_{Q\rightarrow \infty}[V^{(n)}_{3(|Q|+|n|)-m,\mathrm{3body}}]_\mathrm{sphere}=[V^{(n)}_{m,\mathrm{3body}}]_\mathrm{plane}\;.\nonumber
\end{eqnarray}

\begin{figure}[t]
\begin{center}
\includegraphics[width=9.0cm]{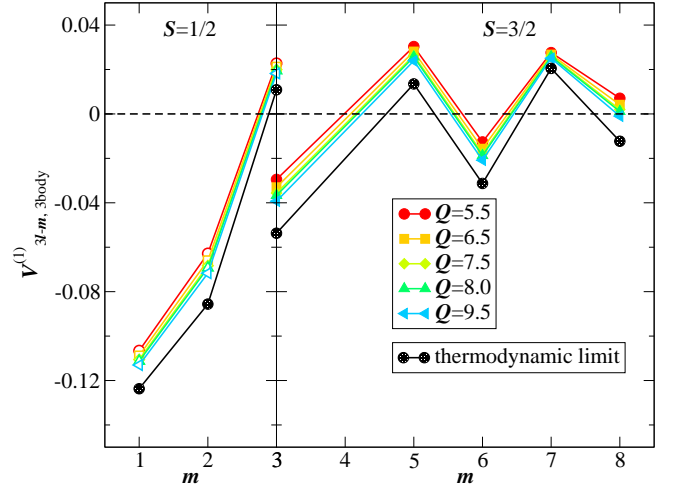}
\caption{(Color online) $V^{(1)}_{3l-m,\mathrm{3body}}$ as a function of relative angular momentum $3l-m$ where $l=|Q|+|n|$ is the single-particle angular momentum at $Q$ in the $n=1$ Landau level.  The pseudopotentials are in units of $e^2/\epsilon l_B$.   
 }  
\label{fig:3bodyvms}
\end{center}
\end{figure}

\begin{figure}[t]
\begin{center}
\includegraphics[width=9.cm]{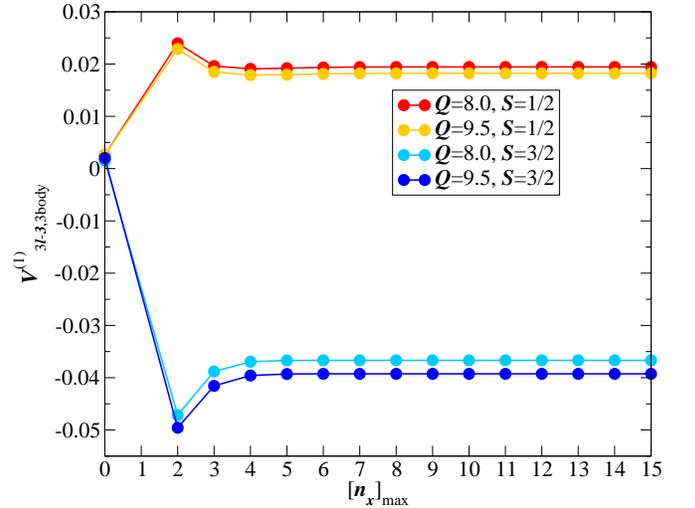}
\caption{(Color online) $V^{(1)}_{3l-3,\mathrm{3body}}$  versus the maximum Landau level used in the finite truncated sum $[n_x]_\mathrm{max}$ in Eq.~(\ref{eq:u3body}) for $Q=9.5$ and 8.5 for total spin $S=1/2$ and 3/2.  Note that here $n_x=1$ is not included in the sum and that the value of $V^{(1)}_{3l-3,\mathrm{3body}}$ is well converged by $[n_x]_\mathrm{max}\geq 6$.  The pseudopotentials are in units of $e^2/\epsilon l_B$. 
 }  
\label{fig:3bodyextrap}
\end{center}
\end{figure}

\subsection{Three-body Landau level mixing pseudopotentials}

In this work we only consider the single-valued three-body pseudopotentials for $m=1$, 2, and 3 for an unpolarized state for total electron spin $S=1/2$ (the spin is in units of $\hbar$) and $m=3$, 5, 6, 7, and 8 for spin-polarized systems with total spin $S=3/2$ (there is no single-valued three electron pseudopotential for $m=4$).  A full analysis of the matrix  three-body pseudopotentials will await further investigation; cf. Refs.~\onlinecite{Simon2007,Davenport2012,Sodemann2013}.

\begin{table*}[t!]
\caption{$V^{(1)}_{3l-m,\mathrm{3body}}(2l)$'s for particular values of $2l=2|Q|+2|n|$'s for $n=1$.  All energies are given in units of $e^2/\epsilon l_B$ where $l_B = R/\sqrt{Q}$.  Note that in the $n=0$ Landau level all the three-body terms vanish due to symmetry.}
\label{table2}
\begin{tabular}{cc|rrrrr|r}
\hline\hline
$m$ & $2S$ & $V^{(1)}_{13-m,\mathrm{3body}}$  & $V^{(1)}_{15-m,\mathrm{3body}}$  & $V^{(1)}_{17-m,\mathrm{3body}}$  &  $V^{(1)}_{18-m,\mathrm{3body}}$ & $V^{(1)}_{21-m,\mathrm{3body}}$ & $\overline{V}^{(1)}_{m,\mathrm{3body}}$ \\
\hline 
1 & 1 & -0.10657 & -0.10882 & -0.11055 & -0.11127  & -0.11303 &  -0.12370  \\
2 & 1 & -0.06282 & -0.06595 & -0.06832 & -0.06931 &  -0.07169  &  -0.08560 \\
3 & 1 &  0.02275 &  0.02117 &  0.01995  & 0.01945  &   0.01823  &   0.01088\\
\hline
3 & 3 & -0.02949  & -0.03299 & -0.03560 &  -0.03668  & -0.03926  &  -0.05371 \\
5 & 3 &  0.03034   & 0.02809 &  0.02637  &  0.02565   &  0.02391 &  0.01345 \\
6 & 3 & -0.01261   & -0.01565 & -0.01785 &  -0.01874  &  -0.02082 &  -0.03132 \\
7 & 3 &  0.02760   & 0.02674  & 0.02607  &  0.02578  & 0.02508  &   0.02045 \\
8 & 3 &  0.00705   & 0.00415 &  0.00199 &   0.00111 &  -0.00100 &   -0.01234 \\
\hline
\hline
\end{tabular}
\end{table*}

We now present  LL mixing pseudopotentials for the spherical geometry for commonly studied system sizes and discuss the three-body pseudopotentials given in Table~\ref{table2} and plotted in Fig.~\ref{fig:3bodyvms}.  Note that each $V_{L,\mathrm{3body}}^{(n)}(\kappa)$ is linear in $\kappa$ and enters the Hamiltonian Eq.~(\ref{eq:LLMham}) as $\kappa V_{L,\mathrm{3body}}^{(n)}$ and only the value of the pseudopotentials are given the table and figure.  All three-body terms  vanish exactly due to symmetry for the lowest LL ($n=0$).  To calculate $V_{L,\mathrm{3body}}^{(n)}$ we solve Eq.~(\ref{eq:3bodyVm}) [and therefore Eq.~(\ref{eq:u3body})] for a finite number of virtual LLs $n_x$, i.e., we truncate the infinite sum.  The careful reader will notice that the $m=6$ pseudopotential in the thermodynamic limit given here (right-most column of Table~\ref{table2}) has the opposite sign than the value appearing originally in Ref.~\onlinecite{Peterson2013}--this was a typo in Ref.~\onlinecite{Peterson2013}, as indicated in a recent erratum~\cite{Peterson2015erratum}.

The final results given in Table~\ref{table2} and plotted in Fig.~\ref{fig:3bodyvms} are the limits of the finite sums as the truncation is taken to infinity.  In general, the three-body terms converge  quickly with $n_x$ and usually are fully converged after including only six LLs in the $n_x$ sum (that is, $\sideset{}{'}\sum_{n_x=-6}^6$ is usually enough to produce convergence)--the convergence is demonstrated for a couple of typical example systems in Fig.~\ref{fig:3bodyextrap}.

The dependence of the pseudopotentials on the spherical radius ($R=l_B\sqrt{Q}$) is relatively mild.  However, there are some interesting nontrivial effects.  For example, the $\overline{V}_{3l-8,\mathrm{3body}}^{(1)}=-0.01234$ (in units of $e^2/\epsilon l_B$) in the thermodynamic limit, however, for moderate finite-sized (and commonly diagonalized) systems in the spherical geometry it is at least a factor of ten smaller (in absolute value) and positive, only achieving a negative value of $-0.00100$ for the $Q=9.5$.  Other nontrivial effects can be seen most clearly in Tanontrivialble~\ref{table2}.

\subsection{Two-body Landau level mixing pseudopotentials}

\begin{figure}[t!]
\begin{center}
\includegraphics[width=9.cm]{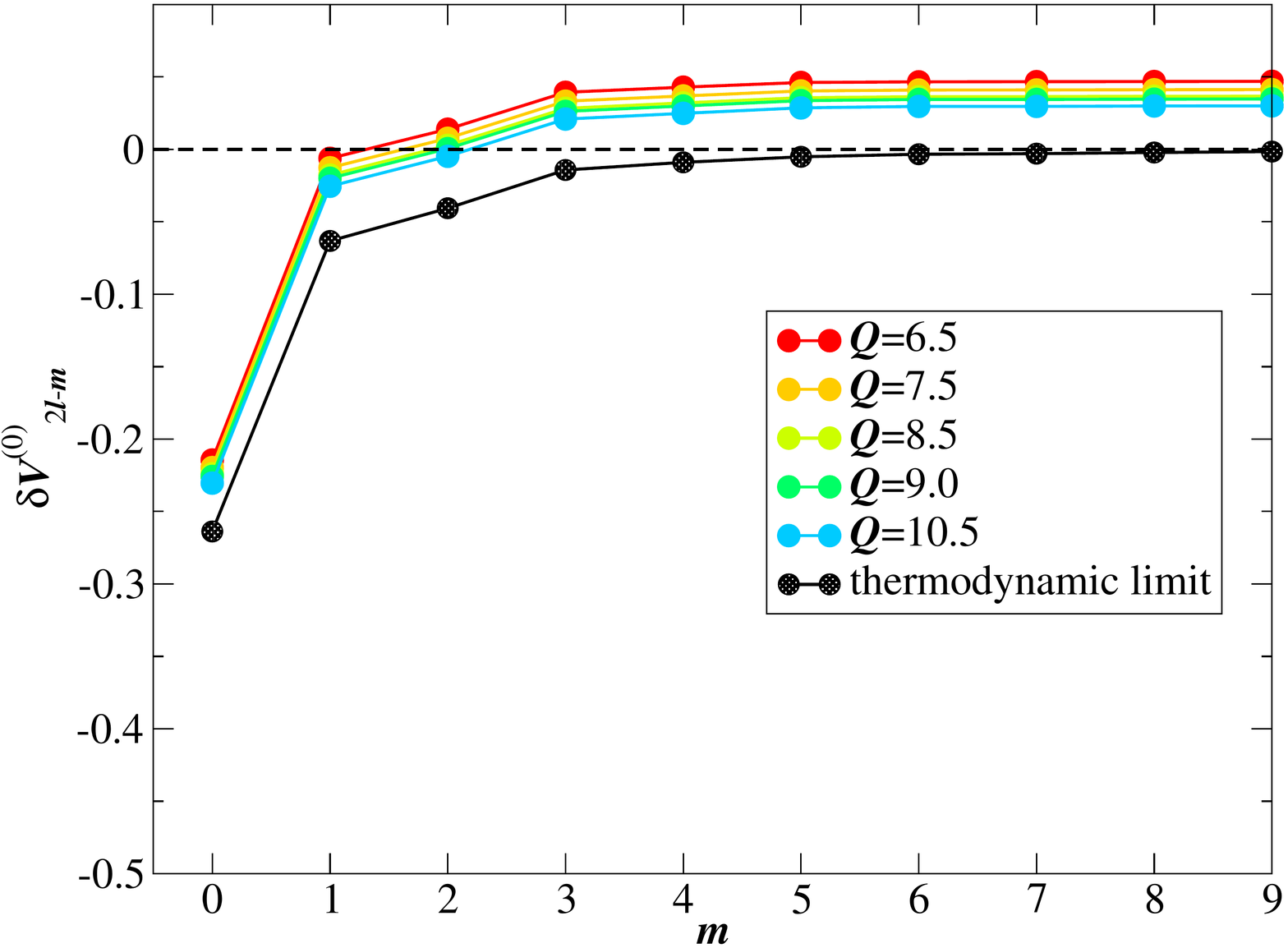}\\
\includegraphics[width=9.cm]{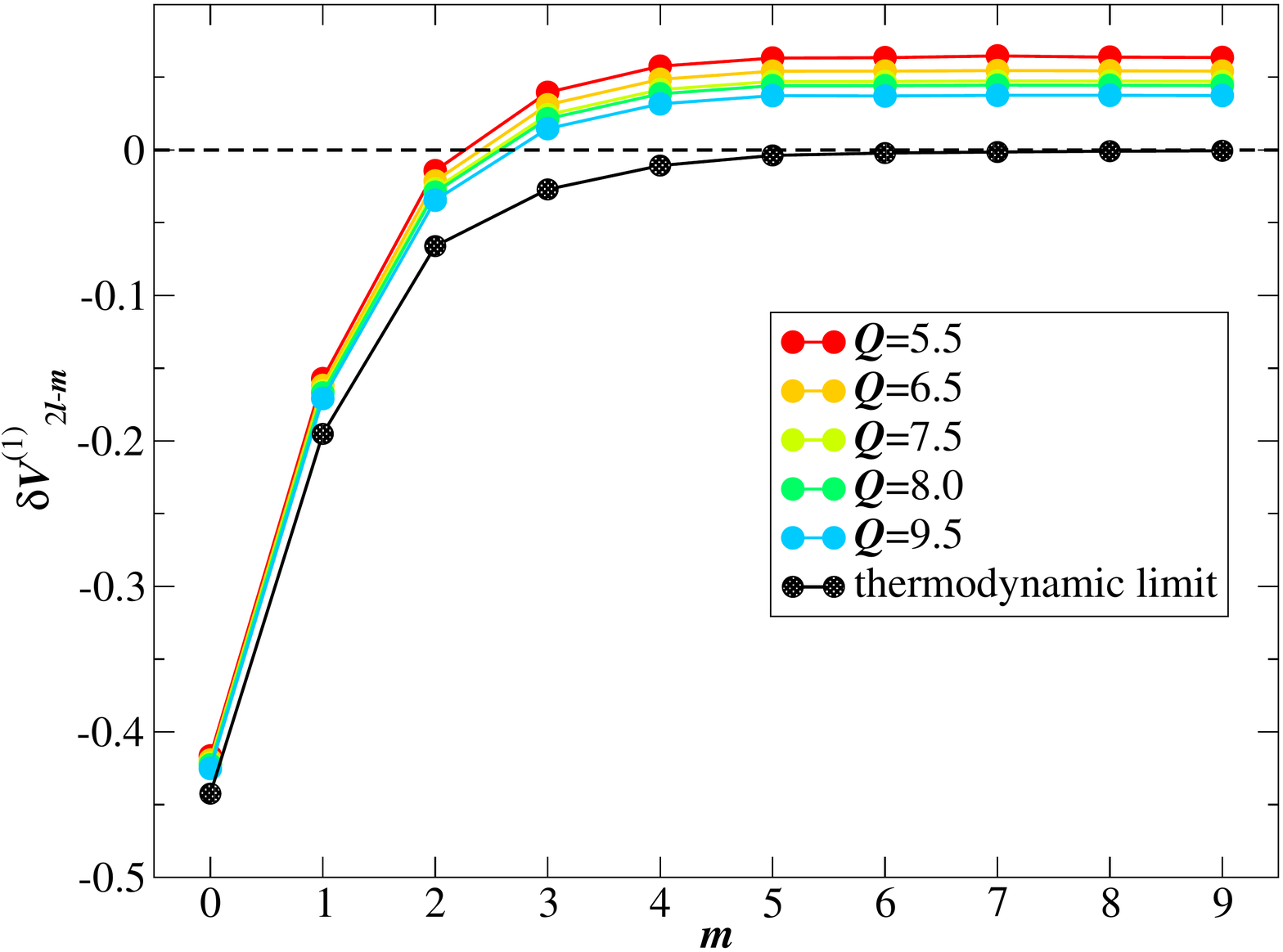}
\caption{(Color online) Two-body pseudopotential corrections due to Landau level mixing ($\delta V^{(n)}_{2l-m,\mathrm{2body}}$) for the lowest ($n=0$, top panel) and first ($n=1$, bottom panel) Landau levels.  $\delta V^{(1)}_{2l-m,\mathrm{2body}}$ is larger in absolute value than $\delta V^{(0)}_{2l-m,\mathrm{2body}}$ and both dramatically decrease with increasing $m$.  The thermodynamic limit ($Q\rightarrow \infty$, denoted $\delta\overline{V}_{m,\mathrm{2body}}$) values are negative for all $m$ for both Landau levels.  However, both Landau level results show nontrivial sign behavior for $m\geq 2$ (for $n=0$) and $m\geq 3$ (for $n=1$).  All energies are given in units of $e^2/\epsilon l_B$.
 }  
\label{fig:n1delta2bodyvms}
\end{center}
\end{figure}

\begin{table*}
\caption{$\delta V^{(n)}_{m,\mathrm{2body}}(2l)$'s for a few values of $2l=2|Q|+2|n|$'s in the $n=0$ and $n=1$ Landau levels, respectively.  All energies are given in units of $e^2/\epsilon l_B$.  }
\label{table3}
\begin{tabular}{c|rrrrr|r}
\hline\hline
$m$ & $\delta V^{(0)}_{13-m,\mathrm{2body}}$  & $\delta V^{(0)}_{15-m,\mathrm{2body}}$  & $\delta V^{(0)}_{17-m,\mathrm{2body}}$  &  $\delta V^{(0)}_{18-m,\mathrm{2body}}$ & $\delta V^{(0)}_{21-m,\mathrm{2body}}$ & $\delta \overline{V}^{(0)}_{m,\mathrm{2body}}$ \\
\hline 
0 & -0.2145  & -0.2197 & -0.2240  & -0.2258 &  -0.2304   &  -0.2638 \\
1 & -0.0062  & -0.0127 & -0.0178  & -0.0200 &  -0.0255   &  -0.0633  \\
2 &  0.0139  &  0.0076  &  0.0027  &  0.0005 &  -0.0048   &  -0.0407  \\
3 &  0.0394  &  0.0332  &  0.0283  &  0.0262 &   0.0209   &  -0.0143  \\
4 &  0.0429  &  0.0368  &  0.0320  &  0.0299 &   0.0248   &  -0.0090  \\
5 &  0.0461  &  0.0403  &  0.0356  &  0.0336 &   0.0286   &  -0.0052 \\
6 &  0.0466  &  0.0409  &  0.0363  &  0.0344 &   0.0296   &  -0.0034 \\
7 &  0.0467 &   0.0410  &  0.0364  &  0.0344 &   0.0296   &  -0.0030 \\
8 &  0.0468 &   0.0411  &  0.0366  &  0.0347 &   0.0299   &  -0.0022 \\
9 &  0.0469  &  0.0413  &  0.0367  &  0.0348 &   0.0300   &  -0.0016 \\
\hline
\hline
$m$ & $\delta V^{(1)}_{13-m,\mathrm{2body}}$  & $\delta V^{(1)}_{15-m,\mathrm{2body}}$  & $\delta V^{(1)}_{17-m,\mathrm{2body}}$  &  $\delta V^{(1)}_{18-m,\mathrm{2body}}$ & $\delta V^{(1)}_{21-m,\mathrm{2body}}$ & $\delta \overline{V}^{(1)}_{m,\mathrm{2body}}$ \\
\hline 
0 & -0.4165 & -0.4194 & -0.4217  & -0.4227 & -0.4252   &  -0.4425 \\
1 & -0.1572 & -0.1619 & -0.1655  & -0.1671 & -0.1709   &  -0.1952 \\
2 & -0.0143 & -0.0214 & -0.0268  & -0.0291 & -0.0345   &  -0.0661  \\
3 &  0.0396 &  0.0310 &  0.0242  &  0.0214 &  0.0147  &  -0.0272  \\
4 &  0.0576 &  0.0485 &  0.0415  &  0.0386 &  0.0316  &  -0.0108  \\
5 &  0.0631 &  0.0541 &  0.0470  &  0.0442 &  0.0373  &  -0.0038 \\
6 &  0.0634 &  0.0542 &  0.0471  &  0.0441 &  0.0371  & -0.0022   \\
7 &  0.0641 &  0.0545 &  0.0474  &  0.0445 &  0.0375  &  -0.0014 \\
8 &  0.0638 &  0.0544 &  0.0473  &  0.0444 &  0.0375  &  -0.0009 \\
9 &  0.0636 &  0.0542 &  0.0471  &  0.0443 &  0.0374  &  -0.0006 \\
\hline
\hline
\end{tabular}
\end{table*}

Finally, we discuss the two-body corrections, $\delta V^{(n)}_{2l-m,\mathrm{2body}}$.  Again we follow the procedure outlined in Ref.~\onlinecite{Peterson2013} and modify the sums and matrix elements for the spherical geometry.  Unlike the three-body terms, the two-body corrections do not vanish for the lowest LL $n=0$.  In Table~\ref{table3} and Fig.~\ref{fig:n1delta2bodyvms} we provide values for $\delta V^{(n)}_{2l-m,\mathrm{2body}}$ for a number of common system sizes for $m=0\ldots9$ in the lowest two LLs.  

Similar to the planar geometry, the values of $\delta V^{(n)}_{2l-m,\mathrm{2body}}$ are, in general, larger in the second $n=1$ LL than they are in the lowest $n=0$ LL.  Furthermore, the values become smaller with increasing $m$ as expected.  In the thermodynamic limit, $\delta\overline{V}^{(n)}_{m,\mathrm{2body}}$ are all negative (as expected).  However, for finite-size systems we find that, for most values of $2l-m$, especially larger $2l-m$ (smaller $m$), the values produce \textit{positive} LL mixing corrections to the bare pseudopotentials and only become progressively smaller and eventually negative for larger systems.  In addition, $\delta V^{(n)}_{2l-m,\mathrm{2body}}$ appear to saturate to a relatively constant, and positive, value by $m\gtrsim 4-5$; this effect is evidentially due to the curvature of the finite sphere.  The qualitative difference between the infinite-system pseudopotentials and finite-size pseudopotentials could have important consequences in exact-diagonalization studies.  

Again, in calculating $\delta V^{(n)}_{2l-m,\mathrm{2body}}$ we truncate the infinite sums 
\begin{eqnarray}
\sideset{}{'}\sum_{n_x=-\infty}^\infty\sideset{}{'}\sum_{n'_x=-\infty}^\infty \rightarrow \sideset{}{'}\sum_{n_x=-[n_x]_\mathrm{max}}^{[n_x]_\mathrm{max}} \sideset{}{'}\sum_{n'_x=-[n_x]_\mathrm{max}}^{[n_x]_\mathrm{max}}\nonumber
\end{eqnarray}
and extrapolate $[n_x]_\mathrm{max}$ to infinity [see Eq. (11) in Ref.~\onlinecite{Peterson2013}].  Interestingly, this extrapolation is simpler in the spherical geometry because, unlike the planar geometry, the sums over intermediate angular momenta are finite.  Because of this we  are able, in this work, to provide more accurate values for $\delta \overline{V}^{(1)}_{m,\mathrm{2body}}$ compared to those given in Ref.~\onlinecite{Peterson2013}.  

\begin{figure}[t!]
\begin{center}
\includegraphics[width=9.cm]{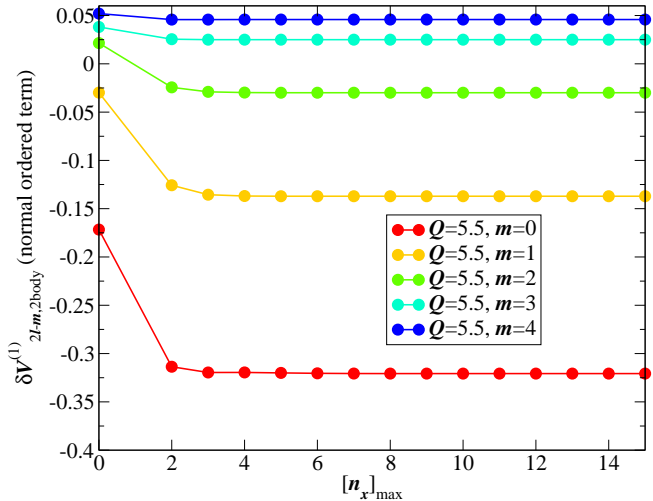}
\caption{(Color online) The normal ordered three-body term contribution to $\delta V^{(1)}_{2l-m,\mathrm{2body}}$ (see Ref.~\onlinecite{Peterson2013}) versus the truncation of the Landau level sum $[n_x]_\mathrm{max}$ for $n=1$ and $Q=5.5$ for $m=0\ldots4$.
 }  
\label{fig:n1delta2bodyNOextrap}
\end{center}
\end{figure}

The reason for these more accurate values is because the two-body LL mixing pseudopotential corrections are composed of two terms: one term is relatively standard and consists of a single loop in a Feynman diagram and are called the ZS, ZS', and BCS terms, respectively, due to their similarity with diagrams from Fermi liquid theory.  The other term arises from a careful normal ordering of the three-body term and does not have a fermion loop; see Ref.~\onlinecite{Peterson2013} for an in-depth discussion.    The terms with one loop contain sums over $n_x$, $n'_x$, $m_x$, and $m'_x$ while the normal ordering term has only $n_x$ and $m_x$ sums.  It is more cumbersome to obtain a reliable extrapolation for the  loop terms, especially in the planar geometry when all sums are infinite.  Furthermore, the loop terms are an order of magnitude smaller, at least, than the terms from normal order.  Hence in Ref.~\onlinecite{Peterson2013} the $n=1$ terms were found by taking the $n=0$ values for the loop terms and using them with the $n=1$ normal order terms.  In this work, the finite nature of the $m_x$ and $m'_x$ sums makes it easy to produce a reliable extrapolation.

\begin{figure}[t!]
\begin{center}
\includegraphics[width=9.cm]{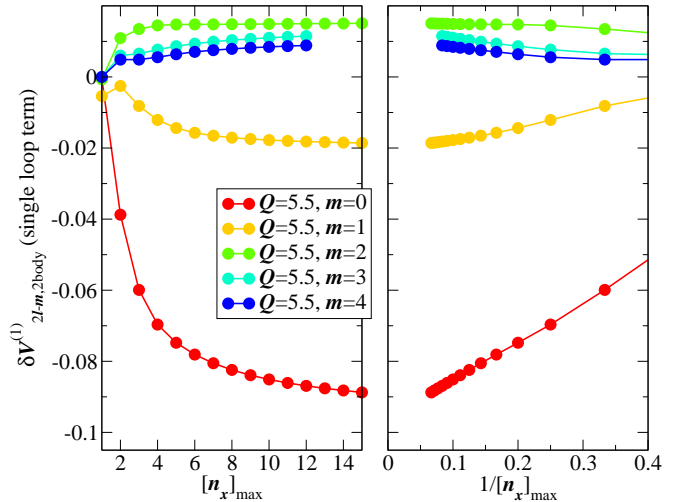}
\caption{(Color online) The loop terms of $\delta V^{(1)}_{2l-m,\mathrm{2body}}$ (ZS, ZS',  and BCS diagram, see text and Ref.~\onlinecite{Peterson2013}) versus the truncation of the loop Landau level sum $[n_x]_\mathrm{max}$  (left panel) and versus $([n_x]_\mathrm{max})^{-1}$ (right panel), respectively, for $n=1$ and $Q=5.5$ for $m=0\ldots4$.
 }  
\label{fig:n1delta2bodyloopextrap}
\end{center}
\end{figure}

The convergence of the two-body term from normal ordering the three-body terms is qualitatively similar to the convergence of the three-body terms, i.e., fast in $[n_x]_\mathrm{max}$ and converged by $[n_x]_\mathrm{max}\sim 6$ (see Fig.~\ref{fig:n1delta2bodyNOextrap} for typical examples).  The term with the fermion loop, however, converges much more slowly.  In Fig.~\ref{fig:n1delta2bodyloopextrap} we plot only the loop terms of $\delta V^{(1)}_{2l-m,\mathrm{2body}}$ versus the truncation of the loop LL sum $[n_x]_\mathrm{max}$ for $Q=5.5$ and $m=0\ldots 4$.  The behavior in this example is typical of other system sizes qualitatively and semiquantitatively.  Clearly the convergence of these terms in $[n_x]_\mathrm{max}$ is much slower than the three-body terms or the two-body terms due to normal ordering of the three-body terms.  In fact, convergence is not achieved until well beyond the inclusion of over 15 LLs in the sums.  In order to determine the convergence in the $[n_x]_\mathrm{max}\rightarrow\infty$ limit we plot the loop terms of $\delta V^{(1)}_{2l-m,\mathrm{2body}}$ versus $([n_x]_\mathrm{max})^{-1}$ and for $[n_x]_\mathrm{max}> 6$, at least, in order to be assured of discounting transient behavior at small $[n_x]_\mathrm{max}$.

\section{Many-body exact diagonalization: spherical versus planar pseudopotentials}
\label{diag}
Before concluding we briefly compare the results of many-body exact diagonalization done by using the spherical versus the infinite planar pseudopotentials as a function of system size.  Specifically we exactly diagonalize  Eq.~(\ref{eq:LLMham}) for the 1/3 filled $n=1$ LL in graphene (recall that the $n=0$ graphene system is identical to that of a GaAs heterostructure in the absence of LL mixing).  Our goal here is not to address a particular physical question, instead, we are estimating the differences in eigenenergies, and potential physical observables, when Eq.~(\ref{eq:LLMham}) is exactly diagonalized by using planar or spherical pseudopotentials.

\begin{figure}[t!]
\begin{center}
\includegraphics[width=9.cm]{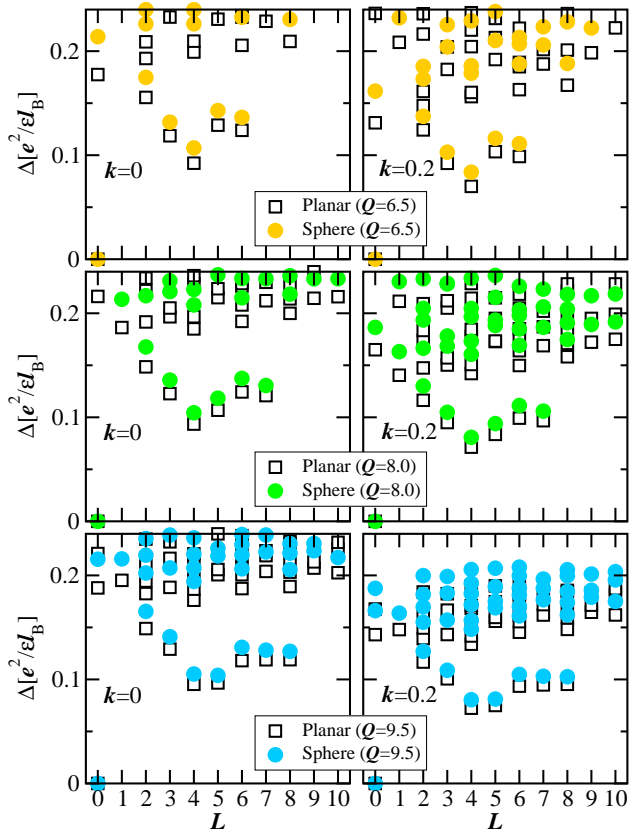}
\caption{(Color online) Energy (measured relative to the ground-state energy) versus total angular $L$ for the 1/3 filled $n=1$ LL for $2l=3(N-1)$ (this relationship corresponds to the 1/3-filled Laughlin state) for $N=6$ ($Q=6.5$), 7 ($Q=8.0$), and 8 ($Q=9.5$).  Circles (squares) represent energies calculated using the spherical (planar) pseudopotentials.  The left panels have zero LL mixing ($\kappa=0$) while the right panels have $\kappa=0.2$.  The results for the spherical pseudopotentials have uniformly larger gaps than those using the planar pseudopotentials, as expected.   As $Q$ (or $N$) increases, the differences in the energies decreases. 
 } 
\label{fig:ed-sphere-vs-plane}
\end{center}
\end{figure}

In Fig.~\ref{fig:ed-sphere-vs-plane} we show eigenenergy spectra, i.e., energy (relative to the ground state) versus total angular momentum $L$, for  filling factor 1/3 in the $n=1$ LL.  We set $2l=3(N-1)$ (corresponding to spherical shift~\cite{wen1990b} for the Laughlin state~\cite{laughlin1983}) projected into the $n=1$ Landau level for $N=6$ ($Q=6.5$), 7 ($Q=8.0$), and 8 ($Q=9.5$) electrons for zero ($\kappa=0$) and finite LL mixing ($\kappa=0.2$), respectively, for illustrative purposes.  (For more details on exact diagonalization in the spherical geometry, please see Refs.~\onlinecite{jainCF,haldane1983}.) Table~\ref{table1} shows that the spherical pseudopotentials are uniformly larger than the planar pseudopotentials at each $m$; thus, it is expected that all energy gaps would be larger when using the spherical pseudopotentials rather than the planar pseudopotentials and, indeed, this is what is observed.  In general, the energy spectrum of the spherical and planar pseudopotentials is qualitatively and quantitatively similar--this remains with or without LL mixing.  We emphasize that if the energy differences between competing FQH (or non-FQH) states at constant filling factor are small, then the the small, but finite, differences in the eigenenergies found when using planar or spherical pseudopotentials could obscure the physics.  As $Q$ increases, the difference between the relative energies decreases as expected because the spherical pseudopotentials extrapolate to the planar ones in the $Q\rightarrow\infty$ limit.  

\begin{figure}[t!]
\begin{center}
\includegraphics[width=9.cm]{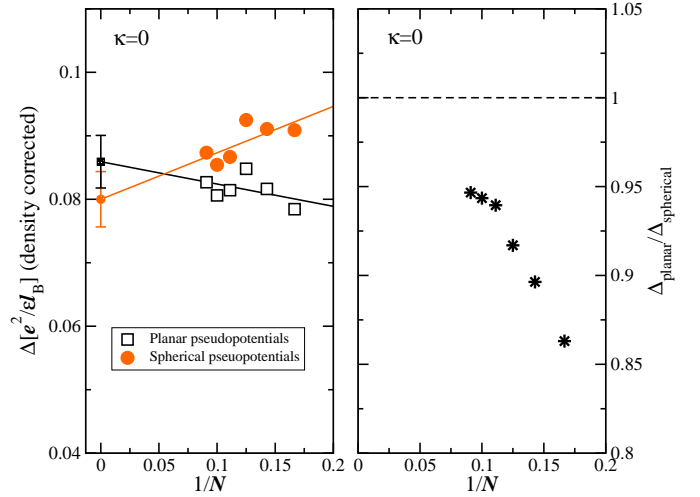}
\caption{(Color online) Energy gap (left panel) versus $1/N$ for $N=6,\ldots, 11$ using the spherical pseudopotentials (circles) and planar pseudopotentials (squares).    The energy has been ``density corrected" by multiplying each raw energy by $\sqrt{\rho_\infty/\rho_N}=\sqrt{2Q\nu/N}$ where $\rho_\infty$ and $\rho_N$ are the electron densities in the thermodynamic limit and for the finite system, respectively.   The gaps using the spherical pseudopotentials are uniformly larger than the gaps calculated using the planar pseudopotentials but the difference between the two decreases with increasing $N$.  The lines are linear extrapolations to the thermodynamic limit with $N=0$ intercepts equal to $0.086\pm 0.004$ $e^2/\epsilon l_\mathrm{B}$ (planar) and $0.080\pm 0.004$ $e^2/\epsilon l_\mathrm{B}$ (sphere), respectively.  The panel shows the ratio of the gaps as a function of $1/N$.  This ratio is approaching unity with increasing $N$, as expected.
 }  
\label{fig:gap-sphere-vs-plane}
\end{center}
\end{figure}

Figure~\ref{fig:gap-sphere-vs-plane} (left panel) displays the energy gap for a far-separated quasiparticle and quasihole pair for $\kappa=0$ versus $1/N$.  This energy gap is the difference between the lowest energy at $L=N$  and the $L=0$ ground state (this is also the smallest energy gap in the spectra for the systems studied).   Again we observe the energy gaps calculated by using the spherical pseudopotentials to be higher than that calculated by using the planar pseudopotentials.  As $N$ increases, i.e., as the thermodynamic limit is approached, the difference in the differently calculated energy gaps  decreases.  A linear extrapolation to the thermodynamic limit yields the same energy gap (when including the standard error) using either pseudopotentials.    To obtain a  quantitative understanding of this difference we plot (right panel of Fig.~\ref{fig:gap-sphere-vs-plane}) the ratio between the gaps calculated by using the planar and spherical pseudopotentials.  For the smallest system considered ($N=6$) the ratio between the energy gaps is $\sim 0.86$ while for the largest system considered ($N=11$) the ratio is $\sim 0.95$.  Thus, the relative error when exactly diagonalizing using spherical versus planar pseudopotentials can be as large as approximately 15\%.

\section{Conclusions}
\label{conc}
In this work we have considered the Landau level problem for massless Dirac fermions in the Haldane spherical geometry commonly used in exact diagonalization studies of the FQHE.  We derived the single-particle eigenstates and eigenenergies  by using spherical cyclotron motion operators $\mathbf{S}$~\cite{Greiter2011}.  These solutions were then used to do two main things:  One was to calculate the Haldane pseudopotentials for the graphene FQHE entirely within the spherical geometry.  This result is important because it has been found that various competing FQH states, e.g., various spin and valley polarizations, are very close in energy and the approach to the thermodynamic limit must be taken with great care to reduce the chance of systematic errors.  In Sec.~\ref{diag} we provided a brief systematic study analyzing the quantitative differences in the many-body spectrum calculated by using the spherical versus planar pseudopotentials.  Second, we fully characterized an effective LL mixing Hamiltonian for graphene specific to the spherical geometry.  (Incidentally, we provided new, more accurate, values for the planar two-body pseudopotential corrections in the $n=1$ LL, i.e., the thermodynamic limit of the spherical values.)  LL mixing is an extremely important effect for the FQHE in graphene, since it cannot be suppressed with the strength of the external magnetic field and must be taken into account in any theoretical treatment that strives toward experimental connections. 
We expect our results (single-particle eigenfunctions and eigenenergies, bare pseudopotentials, and effect LL mixing Hamiltonian) will stimulate further work on the FQHE in graphene and eventually contribute toward the resolution of many of the remaining mysteries.

\textit{Note Added}:  Recently, we learned of Ref.~\onlinecite{Yonaga2016} (and Ref.~\onlinecite{Hasebe2015}) which contained some similar results in the zero-LL-mixing limit. 

\begin{acknowledgements}
This material is based upon work supported by the National Science Foundation under Grant No. DMR-1508290 and we thank the Office of Research and Sponsored Programs at California State University Long Beach and the W. M. Keck Foundation for additional research funding.  We thank Vito Scarola, Peter Raum, and Csaba T\"oke for helpful discussions.  
\end{acknowledgements}

\begin{widetext}
\appendix

\section{Coulomb matrix elements}
\label{app:matrixelement}
We provide the explicit formula for the integral for the Coulomb matrix element [Eq.~(\ref{eq:matrixelement})], which can be found, for example, in Ref.~\onlinecite{jainCF}, but  for the sake of completeness, is reproduced here.  The full form of the monopole harmonics in terms of Haldane spinor coordinates can be written as
\begin{eqnarray}
Y_{Qlm} &=& N_{Qlm}(-1)^{l-m}\sum_{s=0}^{l-m}(-1)^s {l-Q \choose s}{l+Q \choose l-m-s} \bar{v}^{l-Q-s}v^{l-s-m}\bar{u}^s u^{Q+m+s}
\end{eqnarray}
with normalization coefficient 
\begin{eqnarray}
N_{Qlm}=\left[\frac{(2l+1)}{4\pi}\frac{(l-m)!(l+m)!}{(l-Q)!(l+Q)!}\right]^{1/2}\;.
\end{eqnarray}
The Coulomb matrix element is then written as
\begin{eqnarray}
\label{matrixelement}
\langle n_4 m_4, n_3 m_3|V|n_2 m_2,n_1 m_1\rangle^{(Q,n)} &=&
 \int d\mathbf{\Omega}_1d\mathbf{\Omega}_2 \bar{\mathcal{Y}}_{Q_4n_4m_4}(\mathbf\Omega_1)\bar{\mathcal{Y}}_{Q_3n_3m_3}(\mathbf\Omega_2)\frac{1}{|\mathbf{r}_1-\mathbf{r}_2|}\mathcal{Y}_{Q_1n_1m_1}(\mathbf\Omega_2)\mathcal{Y}_{Q_2n_2m_2}(\mathbf\Omega_1)\nonumber\\
&=&\frac{(2l+1)^2(-1)^{Q_4+Q_3-m_4-m_3}}{R}\sum_{l_0=0}^{2l}\sum_{m=-l_0}^{l_0}(-1)^{-m}\nonumber\\
&&\times 
\begin{Bmatrix}
  Q_4+n_4 & l_0 & Q_2 +n_2\\
 m_4 & m & -m_2 
 \end{Bmatrix}
 \begin{Bmatrix}
  Q_4+n_4 & l_0 & Q_2+n_2 \\
 -Q_4 & 0 & Q_2 
 \end{Bmatrix}\nonumber\\
 &&\times
\begin{Bmatrix}
  Q_3+n_3& l_0 & Q_1+n_1 \\
 m_3 & -m & -m_1 
 \end{Bmatrix}
 \begin{Bmatrix}
  Q_3 +n_3& l_0 & Q_1 +n_1\\
 -Q_3 & 0 & Q_1 
 \end{Bmatrix}
\end{eqnarray}
where $\mathcal{Y}_{Qnm}\equiv Y_{Qlm}$ [see Eq.(\ref{y-defn})], $R$ is the radius of the sphere, $|\mathbf{r}_i-\mathbf{r}_j|=\sqrt{2}R|u_1v_2-u_2v_1|$ is the chord distance between two points the sphere, the $\{\cdots\}$ are the Wigner $3$-$j$ symbols, and $\mathbf\Omega = (\theta,\phi)$ are the spherical coordinates of the electrons.  Note that the physical radius of the sphere is set by the value of the monopole strength through $R = l_B\sqrt{Q}$ and single-particle angular momenta $l$ for such a system is constant for all particles and is set by $Q$ and the LL index $n$ through $l=Q+n$.  Thus, each state in the matrix element can have different $Q_i$ and $n_i$ but the combination $l_i = Q_i + n_i$ is constant and $l_i=l$ for all $i$; see the single-particle eigenstates given in Eq.(\ref{eq:eigs-sphere}).  The integral in the first line of Eq.(\ref{matrixelement}) can be calculated by using various identities found in Refs.~\onlinecite{WuYang1976,jainCF}.

\section{Mapping $\mathbf{\Pi}$ to $\mathbf{S}$}
\label{app:map}
Here we provide the derivation of the equalities, $R\Pi_\theta = -S_1$ and $R\Pi_\phi=S_2$.   The identification of  $R\Pi_\phi$ with $S_2$ is trivial.  The differential operators with respect to $\phi$ and $\theta$ can be written in terms of the Haldane spinor coordinates (and their complex conjugates) as
\begin{eqnarray}
\frac{\partial}{\partial\theta} &=& \frac{1}{2}\left(- \bar{v}\frac{\partial}{\partial u} + \bar{u}\frac{\partial}{\partial v} - v\frac{\partial}{\partial \bar{u}} + u\frac{\partial}{\partial \bar{v}}  \right)\\
\frac{\partial}{\partial\phi} &=&  \frac{i}{2}\left(u\frac{\partial}{\partial u} - v\frac{\partial}{\partial v} - \bar{u}\frac{\partial}{\partial \bar{u}} + \bar{v}\frac{\partial}{\partial \bar{v}}\right)\;.
\end{eqnarray}
From these it is clear that 
\begin{eqnarray}
R\Pi_\phi &=& -i\hbar\frac{\partial}{\partial \theta}\\
&=&\hbar \frac{1}{2i}\left(- \bar{v}\frac{\partial}{\partial u} + \bar{u}\frac{\partial}{\partial v} - v\frac{\partial}{\partial \bar{u}} + u\frac{\partial}{\partial \bar{v}}  \right)\\
&=& \frac{1}{2i}(S_+-S_-) \\
&=& S_2\;.
\end{eqnarray}
The identification of $S_1$ with $\Pi_\theta$ is more opaque.  Starting with
\begin{eqnarray}
R\Pi_\theta &=& -i\hbar\frac{1}{\sin(\theta)}\frac{\partial}{\partial\phi} + \cot(\theta)S_3\\
&=&\frac{\hbar}{2\sin\theta}\left(u\frac{\partial}{\partial u} - v\frac{\partial}{\partial v} - \bar{u}\frac{\partial}{\partial\bar{u}} + \bar{v}\frac{\partial}{\partial \bar{v}}\right) + \cot\theta S_3
\end{eqnarray}
we can substitute $1/\sin\theta = \cot\theta + \tan(\theta/2)$  and transform $R\Pi_\theta$ to
\begin{eqnarray}
R\Pi_\theta 
&=&-\hbar\cot\theta\left[\left(\frac{u}{2}\frac{\partial}{\partial u} + \frac{v}{2}\frac{\partial}{\partial v} - \frac{\bar{u}}{2}\frac{\partial}{\partial \bar{u}} - \frac{\bar{v}}{2}\frac{\partial}{\partial \bar{v}}\right) - v\frac{\partial}{\partial v} + \bar{v}\frac{\partial}{\partial\bar{v}}\right]\\
&&-\hbar\tan(\theta/2)\left(\frac{u}{2}\frac{\partial}{\partial u} - \frac{v}{2}\frac{\partial}{\partial v} - \frac{\bar u}{2}\frac{\partial}{\partial \bar{u}} + \frac{\bar{v}}{2}\frac{\partial}{\partial \bar{v}}\right)+\cot\theta S_3\\
&=&-\cot\theta[S_3 - S_3] + \hbar\left(\frac{1}{2}\tan(\theta/2) + \cot\theta\right)\left(v\frac{\partial}{\partial v} - \bar{v}\frac{\partial}{\partial \bar{v}}\right) 
+ \frac{\hbar}{2}\tan(\theta/2)\left(\bar{u}\frac{\partial}{\partial \bar{u}} - u\frac{\partial}{\partial u}\right)\;.
\end{eqnarray}
The first term vanishes and using $\cot(\theta/2)/2 = \cot\theta +  \tan(\theta/2)/2$ we write
\begin{eqnarray}
R\Pi_\theta 
&=& \frac{\hbar}{2}\cot(\theta/2)v\frac{\partial}{\partial v} - \frac{\hbar}{2}\cot(\theta/2)\bar{v}\frac{\partial}{\partial \bar{v}}
+\frac{\hbar}{2} \tan(\theta/2)\bar{u}\frac{\partial}{\partial \bar{u}} - \frac{\hbar}{2} \tan(\theta/2)u\frac{\partial}{\partial u}\;.
\end{eqnarray}
Last, we note $\tan(\theta/2)=v/\bar{u}=\bar{v}/u$ and $\cot(\theta/2)=\bar{u}/v=u/\bar{v}$ to get
\begin{eqnarray}
R\Pi_\theta 
&=& \frac{\hbar}{2}\left(\bar{u}\frac{\partial}{\partial v} - u\frac{\partial}{\partial \bar{v}}
+u\frac{\partial}{\partial \bar{u}} - \bar{v}\frac{\partial}{\partial u}\right)\\
&=&-\frac{1}{2}(S_+ + S_-) = -S_1\;
\end{eqnarray}
completing the derivation.
\end{widetext}


\begin{thebibliography}{43}
\expandafter\ifx\csname natexlab\endcsname\relax\def\natexlab#1{#1}\fi
\expandafter\ifx\csname bibnamefont\endcsname\relax
  \def\bibnamefont#1{#1}\fi
\expandafter\ifx\csname bibfnamefont\endcsname\relax
  \def\bibfnamefont#1{#1}\fi
\expandafter\ifx\csname citenamefont\endcsname\relax
  \def\citenamefont#1{#1}\fi
\expandafter\ifx\csname url\endcsname\relax
  \def\url#1{\texttt{#1}}\fi
\expandafter\ifx\csname urlprefix\endcsname\relax\def\urlprefix{URL }\fi
\providecommand{\bibinfo}[2]{#2}
\providecommand{\eprint}[2][]{\url{#2}}

\bibitem[{\citenamefont{Sarma and Pinczuk}(1996)}]{dasSarma1996}
\bibinfo{author}{\bibfnamefont{S.~D.} \bibnamefont{Sarma}} \bibnamefont{and}
  \bibinfo{author}{\bibfnamefont{A.}~\bibnamefont{Pinczuk}},
  \emph{\bibinfo{title}{Perspectives in Quantum Hall Effects}}
  (\bibinfo{publisher}{Wiley}, \bibinfo{year}{1996}).

\bibitem[{\citenamefont{Jain}(2007)}]{jainCF}
\bibinfo{author}{\bibfnamefont{J.~K.} \bibnamefont{Jain}},
  \emph{\bibinfo{title}{Composite Fermions}} (\bibinfo{publisher}{Cambridge
  University Press}, \bibinfo{year}{2007}).

\bibitem[{\citenamefont{Tsui et~al.}(1982)\citenamefont{Tsui, Stormer, and
  Gossard}}]{tsui1982}
\bibinfo{author}{\bibfnamefont{D.~C.} \bibnamefont{Tsui}},
  \bibinfo{author}{\bibfnamefont{H.~L.} \bibnamefont{Stormer}},
  \bibnamefont{and} \bibinfo{author}{\bibfnamefont{A.~C.}
  \bibnamefont{Gossard}}, \bibinfo{journal}{Phys. Rev. Lett.}
  \textbf{\bibinfo{volume}{48}}, \bibinfo{pages}{1559} (\bibinfo{year}{1982}).

\bibitem[{\citenamefont{Nayak et~al.}(2008)\citenamefont{Nayak, Simon, Stern,
  Freedman, and Sarma}}]{nayak2008}
\bibinfo{author}{\bibfnamefont{C.}~\bibnamefont{Nayak}},
  \bibinfo{author}{\bibfnamefont{S.~H.} \bibnamefont{Simon}},
  \bibinfo{author}{\bibfnamefont{A.}~\bibnamefont{Stern}},
  \bibinfo{author}{\bibfnamefont{M.}~\bibnamefont{Freedman}}, \bibnamefont{and}
  \bibinfo{author}{\bibfnamefont{S.~D.} \bibnamefont{Sarma}},
  \bibinfo{journal}{Rev. Mod. Phys.} \textbf{\bibinfo{volume}{80}},
  \bibinfo{pages}{1083} (\bibinfo{year}{2008}).

\bibitem[{\citenamefont{Moore and Read}(1991)}]{Moore1991}
\bibinfo{author}{\bibfnamefont{G.}~\bibnamefont{Moore}} \bibnamefont{and}
  \bibinfo{author}{\bibfnamefont{N.}~\bibnamefont{Read}},
  \bibinfo{journal}{Nuclear Physics B} \textbf{\bibinfo{volume}{360}},
  \bibinfo{pages}{362 } (\bibinfo{year}{1991}).

\bibitem[{\citenamefont{Das~Sarma et~al.}(2005)\citenamefont{Das~Sarma,
  Freedman, and Nayak}}]{dassarma2005}
\bibinfo{author}{\bibfnamefont{S.}~\bibnamefont{Das~Sarma}},
  \bibinfo{author}{\bibfnamefont{M.}~\bibnamefont{Freedman}}, \bibnamefont{and}
  \bibinfo{author}{\bibfnamefont{C.}~\bibnamefont{Nayak}},
  \bibinfo{journal}{Phys. Rev. Lett.} \textbf{\bibinfo{volume}{94}},
  \bibinfo{pages}{166802} (\bibinfo{year}{2005}).

\bibitem[{\citenamefont{{Novoselov} et~al.}(2004)\citenamefont{{Novoselov},
  {Geim}, {Morozov}, {Jiang}, {Zhang}, {Dubonos}, {Grigorieva}, and
  {Firsov}}}]{Novoselov2004}
\bibinfo{author}{\bibfnamefont{K.~S.} \bibnamefont{{Novoselov}}},
  \bibinfo{author}{\bibfnamefont{A.~K.} \bibnamefont{{Geim}}},
  \bibinfo{author}{\bibfnamefont{S.~V.} \bibnamefont{{Morozov}}},
  \bibinfo{author}{\bibfnamefont{D.}~\bibnamefont{{Jiang}}},
  \bibinfo{author}{\bibfnamefont{Y.}~\bibnamefont{{Zhang}}},
  \bibinfo{author}{\bibfnamefont{S.~V.} \bibnamefont{{Dubonos}}},
  \bibinfo{author}{\bibfnamefont{I.~V.} \bibnamefont{{Grigorieva}}},
  \bibnamefont{and} \bibinfo{author}{\bibfnamefont{A.~A.}
  \bibnamefont{{Firsov}}}, \bibinfo{journal}{Science}
  \textbf{\bibinfo{volume}{306}}, \bibinfo{pages}{666} (\bibinfo{year}{2004}).

\bibitem[{\citenamefont{Bolotin et~al.}(2009)\citenamefont{Bolotin, Ghahari,
  Shulman, Stormer, and Kim}}]{Bolotin2009}
\bibinfo{author}{\bibfnamefont{K.~I.} \bibnamefont{Bolotin}},
  \bibinfo{author}{\bibfnamefont{F.}~\bibnamefont{Ghahari}},
  \bibinfo{author}{\bibfnamefont{M.~D.} \bibnamefont{Shulman}},
  \bibinfo{author}{\bibfnamefont{H.~L.} \bibnamefont{Stormer}},
  \bibnamefont{and} \bibinfo{author}{\bibfnamefont{P.}~\bibnamefont{Kim}},
  \bibinfo{journal}{Nature} \textbf{\bibinfo{volume}{462}},
  \bibinfo{pages}{196} (\bibinfo{year}{2009}).

\bibitem[{\citenamefont{Du et~al.}(2009)\citenamefont{Du, Skachko, Duerr,
  Luican, and Andrei}}]{du2009}
\bibinfo{author}{\bibfnamefont{X.}~\bibnamefont{Du}},
  \bibinfo{author}{\bibfnamefont{I.}~\bibnamefont{Skachko}},
  \bibinfo{author}{\bibfnamefont{F.}~\bibnamefont{Duerr}},
  \bibinfo{author}{\bibfnamefont{A.}~\bibnamefont{Luican}}, \bibnamefont{and}
  \bibinfo{author}{\bibfnamefont{E.~Y.} \bibnamefont{Andrei}},
  \bibinfo{journal}{Nature} \textbf{\bibinfo{volume}{462}},
  \bibinfo{pages}{192} (\bibinfo{year}{2009}).

\bibitem[{\citenamefont{Dean et~al.}(2011)\citenamefont{Dean, Young,
  Cadden-Zimansky, Wang, Ren, Watanabe, Taniguchi, Kim, Hone, and
  Shepard}}]{dean2011}
\bibinfo{author}{\bibfnamefont{C.~R.} \bibnamefont{Dean}},
  \bibinfo{author}{\bibfnamefont{A.~F.} \bibnamefont{Young}},
  \bibinfo{author}{\bibfnamefont{P.}~\bibnamefont{Cadden-Zimansky}},
  \bibinfo{author}{\bibfnamefont{L.}~\bibnamefont{Wang}},
  \bibinfo{author}{\bibfnamefont{H.}~\bibnamefont{Ren}},
  \bibinfo{author}{\bibfnamefont{K.}~\bibnamefont{Watanabe}},
  \bibinfo{author}{\bibfnamefont{T.}~\bibnamefont{Taniguchi}},
  \bibinfo{author}{\bibfnamefont{P.}~\bibnamefont{Kim}},
  \bibinfo{author}{\bibfnamefont{J.}~\bibnamefont{Hone}}, \bibnamefont{and}
  \bibinfo{author}{\bibfnamefont{K.~L.} \bibnamefont{Shepard}},
  \bibinfo{journal}{Nature Phys.} \textbf{\bibinfo{volume}{7}},
  \bibinfo{pages}{693} (\bibinfo{year}{2011}).

\bibitem[{\citenamefont{Feldman et~al.}(2012)\citenamefont{Feldman, Krauss,
  Smet, and Yacoby}}]{feldman2012}
\bibinfo{author}{\bibfnamefont{B.~E.} \bibnamefont{Feldman}},
  \bibinfo{author}{\bibfnamefont{B.}~\bibnamefont{Krauss}},
  \bibinfo{author}{\bibfnamefont{J.~H.} \bibnamefont{Smet}}, \bibnamefont{and}
  \bibinfo{author}{\bibfnamefont{A.}~\bibnamefont{Yacoby}},
  \bibinfo{journal}{Science} \textbf{\bibinfo{volume}{337}},
  \bibinfo{pages}{1196} (\bibinfo{year}{2012}).

\bibitem[{\citenamefont{Feldman et~al.}(2013)\citenamefont{Feldman, Levin,
  Krauss, Abanin, Halperin, Smet, and Yacoby}}]{feldman2013}
\bibinfo{author}{\bibfnamefont{B.~E.} \bibnamefont{Feldman}},
  \bibinfo{author}{\bibfnamefont{A.~J.} \bibnamefont{Levin}},
  \bibinfo{author}{\bibfnamefont{B.}~\bibnamefont{Krauss}},
  \bibinfo{author}{\bibfnamefont{D.~A.} \bibnamefont{Abanin}},
  \bibinfo{author}{\bibfnamefont{B.~I.} \bibnamefont{Halperin}},
  \bibinfo{author}{\bibfnamefont{J.~H.} \bibnamefont{Smet}}, \bibnamefont{and}
  \bibinfo{author}{\bibfnamefont{A.}~\bibnamefont{Yacoby}},
  \bibinfo{journal}{Phys. Rev. Lett.} \textbf{\bibinfo{volume}{111}},
  \bibinfo{pages}{076802} (\bibinfo{year}{2013}).

\bibitem[{\citenamefont{Apalkov and Chakraborty}(2006)}]{Apalkov2006}
\bibinfo{author}{\bibfnamefont{V.~M.} \bibnamefont{Apalkov}} \bibnamefont{and}
  \bibinfo{author}{\bibfnamefont{T.}~\bibnamefont{Chakraborty}},
  \bibinfo{journal}{Phys. Rev. Lett.} \textbf{\bibinfo{volume}{97}},
  \bibinfo{pages}{126801} (\bibinfo{year}{2006}).

\bibitem[{\citenamefont{Goerbig et~al.}(2006)\citenamefont{Goerbig, Moessner,
  and Dou\ifmmode~\mbox{\c{c}}\else \c{c}\fi{}ot}}]{Goerbig2006}
\bibinfo{author}{\bibfnamefont{M.~O.} \bibnamefont{Goerbig}},
  \bibinfo{author}{\bibfnamefont{R.}~\bibnamefont{Moessner}}, \bibnamefont{and}
  \bibinfo{author}{\bibfnamefont{B.}~\bibnamefont{Dou\ifmmode~\mbox{\c{c}}\else
  \c{c}\fi{}ot}}, \bibinfo{journal}{Phys. Rev. B}
  \textbf{\bibinfo{volume}{74}}, \bibinfo{pages}{161407}
  (\bibinfo{year}{2006}).

\bibitem[{\citenamefont{T\ifmmode~\mbox{\H{o}}\else \H{o}\fi{}ke
  et~al.}(2006)\citenamefont{T\ifmmode~\mbox{\H{o}}\else \H{o}\fi{}ke, Lammert,
  Crespi, and Jain}}]{Toke2006}
\bibinfo{author}{\bibfnamefont{C.}~\bibnamefont{T\ifmmode~\mbox{\H{o}}\else
  \H{o}\fi{}ke}}, \bibinfo{author}{\bibfnamefont{P.~E.} \bibnamefont{Lammert}},
  \bibinfo{author}{\bibfnamefont{V.~H.} \bibnamefont{Crespi}},
  \bibnamefont{and} \bibinfo{author}{\bibfnamefont{J.~K.} \bibnamefont{Jain}},
  \bibinfo{journal}{Phys. Rev. B} \textbf{\bibinfo{volume}{74}},
  \bibinfo{pages}{235417} (\bibinfo{year}{2006}).

\bibitem[{\citenamefont{Nomura and MacDonald}(2006)}]{Nomura2006}
\bibinfo{author}{\bibfnamefont{K.}~\bibnamefont{Nomura}} \bibnamefont{and}
  \bibinfo{author}{\bibfnamefont{A.~H.} \bibnamefont{MacDonald}},
  \bibinfo{journal}{Phys. Rev. Lett.} \textbf{\bibinfo{volume}{96}},
  \bibinfo{pages}{256602} (\bibinfo{year}{2006}).

\bibitem[{\citenamefont{Papi{\'c} et~al.}(2009)\citenamefont{Papi{\'c},
  Goerbig, and Regnault}}]{Papic2009}
\bibinfo{author}{\bibfnamefont{Z.}~\bibnamefont{Papi{\'c}}},
  \bibinfo{author}{\bibfnamefont{M.}~\bibnamefont{Goerbig}}, \bibnamefont{and}
  \bibinfo{author}{\bibfnamefont{N.}~\bibnamefont{Regnault}},
  \bibinfo{journal}{Solid State Communications} \textbf{\bibinfo{volume}{149}},
  \bibinfo{pages}{1056 } (\bibinfo{year}{2009}).

\bibitem[{\citenamefont{Sodemann and MacDonald}(2014)}]{Sodemann2014}
\bibinfo{author}{\bibfnamefont{I.}~\bibnamefont{Sodemann}} \bibnamefont{and}
  \bibinfo{author}{\bibfnamefont{A.~H.} \bibnamefont{MacDonald}},
  \bibinfo{journal}{Phys. Rev. Lett.} \textbf{\bibinfo{volume}{112}},
  \bibinfo{pages}{126804} (\bibinfo{year}{2014}).

\bibitem[{\citenamefont{Wu et~al.}(2014)\citenamefont{Wu, Sodemann, Araki,
  MacDonald, and Jolicoeur}}]{Wu2014}
\bibinfo{author}{\bibfnamefont{F.}~\bibnamefont{Wu}},
  \bibinfo{author}{\bibfnamefont{I.}~\bibnamefont{Sodemann}},
  \bibinfo{author}{\bibfnamefont{Y.}~\bibnamefont{Araki}},
  \bibinfo{author}{\bibfnamefont{A.~H.} \bibnamefont{MacDonald}},
  \bibnamefont{and}
  \bibinfo{author}{\bibfnamefont{T.}~\bibnamefont{Jolicoeur}},
  \bibinfo{journal}{Phys. Rev. B} \textbf{\bibinfo{volume}{90}},
  \bibinfo{pages}{235432} (\bibinfo{year}{2014}).

\bibitem[{\citenamefont{Balram et~al.}(2015)\citenamefont{Balram,
  T\ifmmode~\mbox{\H{o}}\else \H{o}\fi{}ke, W\'ojs, and Jain}}]{Balram2015}
\bibinfo{author}{\bibfnamefont{A.~C.} \bibnamefont{Balram}},
  \bibinfo{author}{\bibfnamefont{C.}~\bibnamefont{T\ifmmode~\mbox{\H{o}}\else
  \H{o}\fi{}ke}}, \bibinfo{author}{\bibfnamefont{A.}~\bibnamefont{W\'ojs}},
  \bibnamefont{and} \bibinfo{author}{\bibfnamefont{J.~K.} \bibnamefont{Jain}},
  \bibinfo{journal}{Phys. Rev. B} \textbf{\bibinfo{volume}{92}},
  \bibinfo{pages}{205120} (\bibinfo{year}{2015}).

\bibitem[{\citenamefont{Peterson and Nayak}(2014)}]{Peterson2014a}
\bibinfo{author}{\bibfnamefont{M.~R.} \bibnamefont{Peterson}} \bibnamefont{and}
  \bibinfo{author}{\bibfnamefont{C.}~\bibnamefont{Nayak}},
  \bibinfo{journal}{Phys. Rev. Lett.} \textbf{\bibinfo{volume}{113}},
  \bibinfo{pages}{086401} (\bibinfo{year}{2014}).

\bibitem[{\citenamefont{W\'ojs et~al.}(2011)\citenamefont{W\'ojs, M\"oller, and
  Cooper}}]{Wojs2011}
\bibinfo{author}{\bibfnamefont{A.}~\bibnamefont{W\'ojs}},
  \bibinfo{author}{\bibfnamefont{G.}~\bibnamefont{M\"oller}}, \bibnamefont{and}
  \bibinfo{author}{\bibfnamefont{N.~R.} \bibnamefont{Cooper}},
  \bibinfo{journal}{Acta Physica Polonica A} \textbf{\bibinfo{volume}{119}},
  \bibinfo{pages}{592} (\bibinfo{year}{2011}).

\bibitem[{\citenamefont{Kazama et~al.}(1977)\citenamefont{Kazama, Yang, and
  Goldhaber}}]{Kazama1977}
\bibinfo{author}{\bibfnamefont{Y.}~\bibnamefont{Kazama}},
  \bibinfo{author}{\bibfnamefont{C.~N.} \bibnamefont{Yang}}, \bibnamefont{and}
  \bibinfo{author}{\bibfnamefont{A.~S.} \bibnamefont{Goldhaber}},
  \bibinfo{journal}{Phys. Rev. D} \textbf{\bibinfo{volume}{15}},
  \bibinfo{pages}{2287} (\bibinfo{year}{1977}).

\bibitem[{\citenamefont{Torres~del Castillo and
  Cortes-Cuautli}(1997)}]{Castillo1997}
\bibinfo{author}{\bibfnamefont{G.~F.} \bibnamefont{Torres~del Castillo}}
  \bibnamefont{and} \bibinfo{author}{\bibfnamefont{L.~C.}
  \bibnamefont{Cortes-Cuautli}}, \bibinfo{journal}{J. Math. Phys.}
  \textbf{\bibinfo{volume}{38}}, \bibinfo{pages}{2996} (\bibinfo{year}{1997}).

\bibitem[{\citenamefont{Schliemann}(2008)}]{Schliemann2008}
\bibinfo{author}{\bibfnamefont{J.}~\bibnamefont{Schliemann}},
  \bibinfo{journal}{Phys. Rev. B} \textbf{\bibinfo{volume}{78}},
  \bibinfo{pages}{195426} (\bibinfo{year}{2008}).

\bibitem[{\citenamefont{Newman and Penrose}(1962)}]{Newman1962}
\bibinfo{author}{\bibfnamefont{E.}~\bibnamefont{Newman}} \bibnamefont{and}
  \bibinfo{author}{\bibfnamefont{R.}~\bibnamefont{Penrose}},
  \bibinfo{journal}{J. Math. Phys.} \textbf{\bibinfo{volume}{3}},
  \bibinfo{pages}{566} (\bibinfo{year}{1962}).

\bibitem[{\citenamefont{Dray}(1985)}]{Dray1985}
\bibinfo{author}{\bibfnamefont{T.}~\bibnamefont{Dray}}, \bibinfo{journal}{J.
  Math. Phys.} \textbf{\bibinfo{volume}{26}}, \bibinfo{pages}{1030}
  (\bibinfo{year}{1985}).

\bibitem[{\citenamefont{Jellal}(2008)}]{Jellal2008}
\bibinfo{author}{\bibfnamefont{A.}~\bibnamefont{Jellal}},
  \bibinfo{journal}{Nuclear Physics B} \textbf{\bibinfo{volume}{804}},
  \bibinfo{pages}{361 } (\bibinfo{year}{2008}).

\bibitem[{\citenamefont{Greiter}(2011)}]{Greiter2011}
\bibinfo{author}{\bibfnamefont{M.}~\bibnamefont{Greiter}},
  \bibinfo{journal}{Phys. Rev. B} \textbf{\bibinfo{volume}{83}},
  \bibinfo{pages}{115129} (\bibinfo{year}{2011}).

\bibitem[{\citenamefont{Zaletel et~al.}(2015)\citenamefont{Zaletel, Mong,
  Pollmann, and Rezayi}}]{ZaletelPRB2015}
\bibinfo{author}{\bibfnamefont{M.~P.} \bibnamefont{Zaletel}},
  \bibinfo{author}{\bibfnamefont{R.~S.~K.} \bibnamefont{Mong}},
  \bibinfo{author}{\bibfnamefont{F.}~\bibnamefont{Pollmann}}, \bibnamefont{and}
  \bibinfo{author}{\bibfnamefont{E.~H.} \bibnamefont{Rezayi}},
  \bibinfo{journal}{Phys. Rev. B} \textbf{\bibinfo{volume}{91}},
  \bibinfo{pages}{045115} (\bibinfo{year}{2015}).

\bibitem[{\citenamefont{Haldane}(1983)}]{haldane1983}
\bibinfo{author}{\bibfnamefont{F.~D.~M.} \bibnamefont{Haldane}},
  \bibinfo{journal}{Phys. Rev. Lett.} \textbf{\bibinfo{volume}{51}},
  \bibinfo{pages}{605} (\bibinfo{year}{1983}).

\bibitem[{\citenamefont{Wu and Yang}(1976)}]{WuYang1976}
\bibinfo{author}{\bibfnamefont{T.~T.} \bibnamefont{Wu}} \bibnamefont{and}
  \bibinfo{author}{\bibfnamefont{C.~N.} \bibnamefont{Yang}},
  \bibinfo{journal}{Nuclear Physics B} \textbf{\bibinfo{volume}{107}},
  \bibinfo{pages}{365 } (\bibinfo{year}{1976}).

\bibitem[{\citenamefont{T\ifmmode~\mbox{\H{o}}\else \H{o}\fi{}ke and
  Jain}(2007)}]{Toke2007}
\bibinfo{author}{\bibfnamefont{C.}~\bibnamefont{T\ifmmode~\mbox{\H{o}}\else
  \H{o}\fi{}ke}} \bibnamefont{and} \bibinfo{author}{\bibfnamefont{J.~K.}
  \bibnamefont{Jain}}, \bibinfo{journal}{Phys. Rev. B}
  \textbf{\bibinfo{volume}{75}}, \bibinfo{pages}{245440}
  (\bibinfo{year}{2007}).

\bibitem[{\citenamefont{Laughlin}(1983)}]{laughlin1983}
\bibinfo{author}{\bibfnamefont{R.~B.} \bibnamefont{Laughlin}},
  \bibinfo{journal}{Phys. Rev. Lett.} \textbf{\bibinfo{volume}{50}},
  \bibinfo{pages}{1395} (\bibinfo{year}{1983}).

\bibitem[{\citenamefont{Bishara and Nayak}(2009)}]{Bishara2009}
\bibinfo{author}{\bibfnamefont{W.}~\bibnamefont{Bishara}} \bibnamefont{and}
  \bibinfo{author}{\bibfnamefont{C.}~\bibnamefont{Nayak}},
  \bibinfo{journal}{Phys. Rev. B} \textbf{\bibinfo{volume}{80}},
  \bibinfo{pages}{121302} (\bibinfo{year}{2009}).

\bibitem[{\citenamefont{Peterson and Nayak}(2013)}]{Peterson2013}
\bibinfo{author}{\bibfnamefont{M.~R.} \bibnamefont{Peterson}} \bibnamefont{and}
  \bibinfo{author}{\bibfnamefont{C.}~\bibnamefont{Nayak}},
  \bibinfo{journal}{Phys. Rev. B} \textbf{\bibinfo{volume}{87}},
  \bibinfo{pages}{245129} (\bibinfo{year}{2013}).

\bibitem[{\citenamefont{Simon et~al.}(2007)\citenamefont{Simon, Rezayi, and
  Cooper}}]{Simon2007}
\bibinfo{author}{\bibfnamefont{S.~H.} \bibnamefont{Simon}},
  \bibinfo{author}{\bibfnamefont{E.~H.} \bibnamefont{Rezayi}},
  \bibnamefont{and} \bibinfo{author}{\bibfnamefont{N.~R.}
  \bibnamefont{Cooper}}, \bibinfo{journal}{Phys. Rev. B}
  \textbf{\bibinfo{volume}{75}}, \bibinfo{pages}{195306}
  (\bibinfo{year}{2007}).

\bibitem[{\citenamefont{Davenport and Simon}(2012)}]{Davenport2012}
\bibinfo{author}{\bibfnamefont{S.~C.} \bibnamefont{Davenport}}
  \bibnamefont{and} \bibinfo{author}{\bibfnamefont{S.~H.} \bibnamefont{Simon}},
  \bibinfo{journal}{Phys. Rev. B} \textbf{\bibinfo{volume}{85}},
  \bibinfo{pages}{075430} (\bibinfo{year}{2012}).

\bibitem[{\citenamefont{Sodemann and MacDonald}(2013)}]{Sodemann2013}
\bibinfo{author}{\bibfnamefont{I.}~\bibnamefont{Sodemann}} \bibnamefont{and}
  \bibinfo{author}{\bibfnamefont{A.~H.} \bibnamefont{MacDonald}},
  \bibinfo{journal}{Phys. Rev. B} \textbf{\bibinfo{volume}{87}},
  \bibinfo{pages}{245425} (\bibinfo{year}{2013}).

\bibitem[{\citenamefont{Peterson and Nayak}(2015)}]{Peterson2015erratum}
\bibinfo{author}{\bibfnamefont{M.~R.} \bibnamefont{Peterson}} \bibnamefont{and}
  \bibinfo{author}{\bibfnamefont{C.}~\bibnamefont{Nayak}},
  \bibinfo{journal}{Phys. Rev. B} \textbf{\bibinfo{volume}{92}},
  \bibinfo{pages}{159902} (\bibinfo{year}{2015}).

\bibitem[{\citenamefont{Wen and Niu}(1990)}]{wen1990b}
\bibinfo{author}{\bibfnamefont{X.~G.} \bibnamefont{Wen}} \bibnamefont{and}
  \bibinfo{author}{\bibfnamefont{Q.}~\bibnamefont{Niu}},
  \bibinfo{journal}{Phys. Rev. B} \textbf{\bibinfo{volume}{41}},
  \bibinfo{pages}{9377} (\bibinfo{year}{1990}).

\bibitem[{\citenamefont{{Yonaga} et~al.}(2016)\citenamefont{{Yonaga}, {Hasebe},
  and {Shibata}}}]{Yonaga2016}
\bibinfo{author}{\bibfnamefont{K.}~\bibnamefont{{Yonaga}}},
  \bibinfo{author}{\bibfnamefont{K.}~\bibnamefont{{Hasebe}}}, \bibnamefont{and}
  \bibinfo{author}{\bibfnamefont{N.}~\bibnamefont{{Shibata}}},
  \bibinfo{journal}{Phys. Rev. B} \textbf{\bibinfo{volume}{93}}, 
  \bibinfo{pages}{235122} (\bibinfo{year}{2016}).

\bibitem[{\citenamefont{{Hasebe}}(2015)}]{Hasebe2015}
\bibinfo{author}{\bibfnamefont{K.}~\bibnamefont{{Hasebe}}},
  \bibinfo{journal}{ArXiv e-prints}  (\bibinfo{year}{2015}),
  \eprint{1511.04681}.

\end{thebibliography}
\end{document}